\documentclass[prl,aps,twocolumn,superscriptaddress,fleqn]{revtex4-2}

\usepackage{
	graphicx,
	subfigure,
	placeins
}
\graphicspath{{Figures/}}

\usepackage{graphicx,subfigure,placeins,wrapfig}

\usepackage[english]{babel}
\usepackage{blindtext}

\usepackage[utf8]{inputenc}
\usepackage[T1]{fontenc}
\usepackage{xcolor,booktabs,color,lipsum,makecell,microtype,fix-cm}
\usepackage[normalem]{ulem}
\usepackage{tabularx}
\usepackage{ctable}

\usepackage{bm, amsmath,float,amssymb,upgreek,physics,
	mhchem}
\usepackage[exponent-product=\cdot]{siunitx}

\usepackage{nicefrac} 			

\usepackage{chngcntr,printlen}
\usepackage[unicode=true,colorlinks=true,citecolor=blue,urlcolor=blue]{hyperref}

\begin{document}

\title{Terahertz radiation driven nonlinear transport phenomena in two-dimensional tellurene}

\author{E. Mönch}
\affiliation{Physics Department, University of Regensburg, 93040 Regensburg, Germany}

\author{M. D. Moldavskaya}
\affiliation{Physics Department, University of Regensburg, 93040 Regensburg, Germany}

\author{L. E. Golub}
\affiliation{Physics Department, University of Regensburg, 93040 Regensburg, Germany}

\author{V. V. Bel'kov}
\affiliation{Physics Department, University of Regensburg, 93040 Regensburg, Germany}

\author{J. Wunderlich}
\affiliation{Physics Department, University of Regensburg, 93040 Regensburg, Germany}
\affiliation{Institute of Physics, Czech Academy of Sciences, Cukrovarnická 10, 162 00 Praha 6, Czech
	Republic}

\author{D.~Weiss}
\affiliation{Physics Department, University of Regensburg, 93040 Regensburg, Germany}

\author{J. Gumenjuk-Sichevska}
\affiliation{Johannes Gutenberg-University Mainz, D-55128 Mainz, Germany}
\affiliation{V. Lashkaryov Institute of Semiconductor Physics, National Academy of Science, 03028, Kyiv, Ukraine}

\author{Chang Niu}
\affiliation{Elmore Family School of Electrical and	Computer Engineering, Purdue University, West Lafayette, Indiana 47907, United States}
\affiliation{Birck Nanotechnology Center, Purdue University, West Lafayette, Indiana 47907, United
	States}

\author{Peide D. Ye}
\affiliation{Elmore Family School of Electrical and	Computer Engineering, Purdue University, West Lafayette, Indiana 47907, United States}
\affiliation{Birck Nanotechnology Center, Purdue University, West Lafayette, Indiana 47907, United
	States}

\author{S. D. Ganichev}
\affiliation{Physics Department, University of Regensburg, 93040 Regensburg, Germany}
\affiliation{CENTERA Labs, Institute of High Pressure Physics, PAS, 01 - 142 Warsaw, Poland}
\email{sergey.ganichev@ur.de}

		\begin{abstract}
					Nonlinear electron transport induced by polarized terahertz radiation is studied in two-dimensional tellurene at room temperature. A direct current, quadratic in the radiation's electric field, is observed. Contributions sensitive to radiation helicity, polarization orientation as well as polarization independent current are found. We show that these contributions can be modified by the magnitude of the external gate potential. We demonstrate that this terahertz-driven electric current arises from the Berry curvature dipole and the side-jump microscopic mechanisms.
		\end{abstract}

	\maketitle 



\section{Introduction}

Tellurene, the two-dimensional (2D) manifestation of tellurium, has been recently synthesized, adding a fascinating new member to the family of elemental van der Waals materials. Its discovery marks a significant step forward in expanding the capabilities and applications of 2D materials~\cite{Wu2018b,Wu2018a,Shi2020,Qiu2022,Zha2024}. Tellurene exhibits a range of exceptional properties, including remarkable stability and catalytic activity~\cite{Wu2017}, tunable carrier density and type via electric gating~\cite{Wu2018b}, strain-adjustable bandgap~\cite{Niu2023b}, efficient piezoelectric behavior~\cite{Apte2021,Rao2022}, anisotropic photoresponse~\cite{Gao2019}, low thermal conductivity, and high carrier mobility---reaching about thousand cm$^2$/Vs at room temperature~\cite{Wang2018}. With its outstanding properties, tellurene is destined for use in cutting-edge devices such as photodetectors~\cite{Amani2018}, modulators~\cite{Wu2018a}, saturable absorbers, mode-locking lasers~\cite{Zhang2020}, and field-effect transistors~\cite{Wang2018}. Beyond its practical applications, tellurene has also demonstrated fundamental phenomena, including the quantum Hall effect~\cite{Qiu2020,Niu2021}, spin Hall effect~\cite{Sachdeva2023}, weak antilocalization~\cite{Niu2020}, and nonlinear magnetoresistance~\cite{Niu2023,Ma2024}.

While linear transport phenomena in tellurene remain the primary focus of current research, nonlinear effects---proven to be powerful tools for probing nonequilibrium electronic processes and revealing fundamental properties---are far less explored. Recently, a significant breakthrough was achieved with the detection of a giant nonlinear Hall effect (NLH) in tellurene, highlighting its remarkable nonlinear response~\cite{Cheng2024, Kim2024}. The direct current (dc) generated in response to a static electric field $\bm E$  is expressed by the following relation:  
\begin{equation}
	j^\alpha = \sigma_{\alpha\mu}^{(1)}E_\mu + \sigma_{\alpha\mu\nu}^{(2)}E_\mu E_\nu \,.
	\label{intro}
\end{equation}
Here $\hat{\bm \sigma}^{(1)}$ and $\hat{\bm \sigma}^{(2)}$ are the first (linear conductivity) and second (nonlinear conductivity) order in electric field dc conductivity tensors, respectively, and $\alpha$, $\mu$, and $\nu$ 
run over in-plane Cartesian coordinates. This second-order effect comprises both the transverse NLH $j^y=\sigma_{yxx}^{(2)}E_x^2$ and, the less-studied nonlinear longitudinal current (NLL) $j^x =\sigma_{xxx}^{(2)}E_x^2$. 

The NLH was first proposed in Ref.~\cite{Sodemann2015}, where it was shown that in metals lacking inversion symmetry a Hall-like current due to the Berry curvature dipole in momentum space can emerge. This effect has since been extended to other noncentrosymmetric materials, including semiconductors, attracting increasing attention~\cite{Ma2018a,Du2021,Du2021a,Zhang2021a, Ortix2021,Huang2022a,Yar2022, Gao2023,Huang2023,Li2023,Bandyopadhyay2024,Lee2024}. Investigating these phenomena at high frequencies, comparable to the momentum relaxation rate, promises to reveal new effects and offer a powerful tool for studying nonlinear transport.

In this letter, we present the observation and investigation of nonlinear electron transport phenomena in tellurene, driven by terahertz (THz) radiation at room temperature. In our high-frequency experiments, the resulting dc current arises as a second-order response to the ac electric field $\bm E_\omega(t) = \bm E \exp(-i\omega t)+\bm E^* \exp(i\omega t)$, where $\omega$  is the driving frequency. This current is given by  
\begin{align}
	\label{phen_ac}
	j^\alpha &= \sigma_{\alpha\mu\nu}^{(2)}(\omega)E_\mu E^*_\nu  \nonumber\\
	&= \chi_{\alpha\mu\nu}(E_\mu E_\nu^*+E_\mu^* E_\nu) +  \gamma_{\alpha\mu}i[\bm E \times \bm E^*]_\mu\,,
\end{align}
where $\hat{\bm \chi}(\omega)$ and $\hat{\bm \gamma}(\omega)$  are the third and second rank tensors, respectively. While the first term is a high-frequency counterpart of the second one in Eq.~\eqref{intro}, the  dc current described by the second one has an opposite sign for  clockwise and counterclockwise rotating fields and not detectable in standard dc transport experiments. The latter one has been previously studied in bulk tellurium for both direct optical intersubband as well as indirect intraband transitions~\cite{Asnin1978, Tsirkin2018,Moldavskaya2023}
and is called the circular photogalvanic effect (CPGE)~\cite{Ivchenko2005,Ganichev2005,Ivchenko2018}. The first term on the right-hand side of Eq.~\eqref{phen_ac} represents the linear photogalvanic effect (LPGE)~\cite{Ivchenko2005,Ganichev2005,Ivchenko2018}, a dc current generated by linearly polarized radiation that is sensitive to the orientation of the in-plane THz electric field. As detailed below, we observe both types of currents in tellurene under THz illumination. We attribute the THz-induced electric currents to two key microscopic mechanisms, arising from the low spatial symmetry of tellurene: the intrinsic contribution from the Berry curvature dipole and an extrinsic contribution caused by electron wave packet side-jumps during momentum scattering.

\begin{figure}[t]
	\centering
	\includegraphics[width=\linewidth]{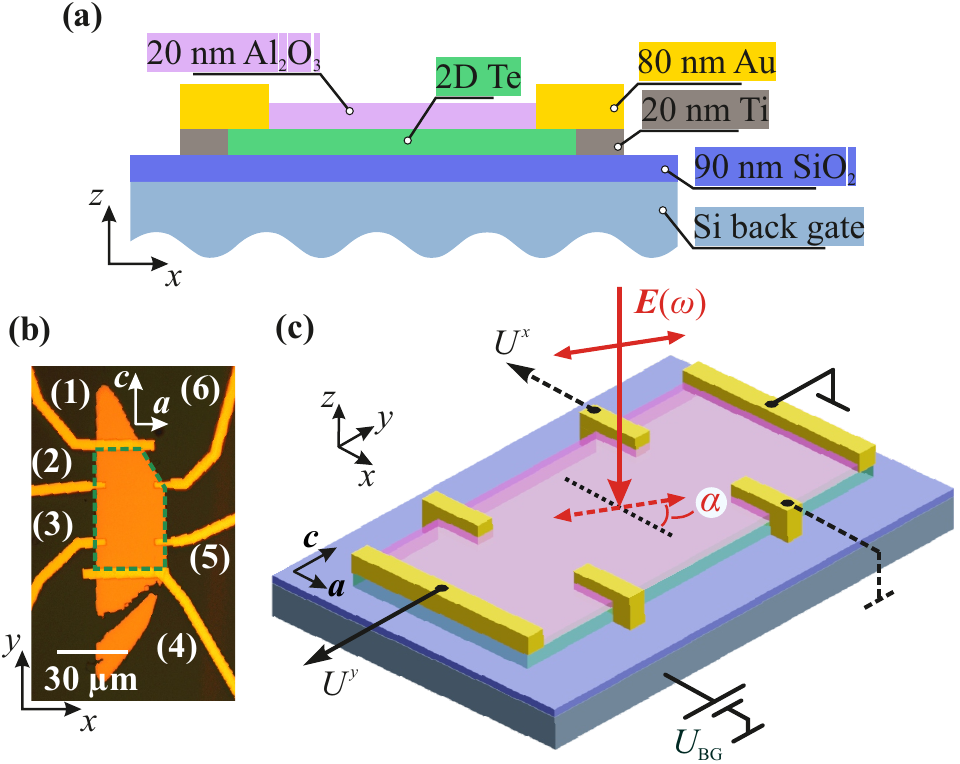}
	\caption{Panel (a) shows a typical cross-section of the 2D Te devices under investigation. Panel (b) shows an optical micrograph of sample \#A and the corresponding contact numbering. The green dashed pentagon highlights the Hall bar. Panel (c) shows a schematic of the photovoltage measurement shown as an example for linearly polarized radiation. The dashed line marks the initial polarization state, i.e., $\alpha = 0$ and $\varphi = 0$. To change the concentration and carrier type, a voltage $U_{\rm BG}$ is applied to the back gate.
	}
	\label{fig1}
\end{figure}

\begin{figure*}[t] 
	\centering
	\includegraphics[width=0.9\linewidth]{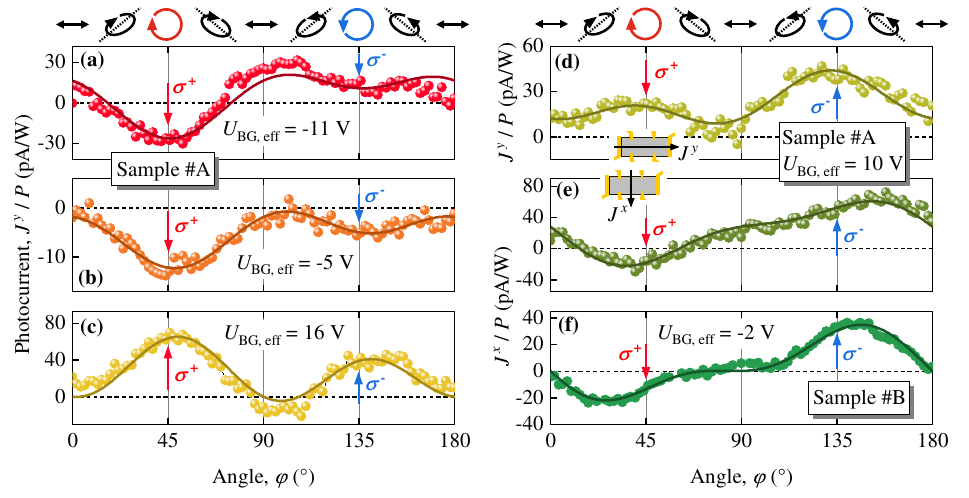}
	\caption{Dependencies of the photocurrent (colored circles) on the rotation angle of the $\lambda/4$ wave plate. The data are shown for several examples of effective back gate voltages ranging from hole ($U_{\rm BG, eff} < 0$) to electron conductivity ($U_{\rm BG, eff} > 0$).  The photoresponse is normalized to the laser power incident on the sample. Panels (a)-(c) show normalized photocurrent, $J^y/P$, obtained for sample \#A. Panels (d) and (e) show $J^y/P$ and $J^x/P$ measured for $U_{\rm BG, eff} = 10$~V measured in sample \#A. Panel (f) shows the data for sample \#B. The red and blue arrows in each panel mark the right-handed ($\sigma^+$) and left-handed ($\sigma^-$) circular polarization states, respectively. At the top, the polarization ellipses corresponding to some angles $\varphi$ are shown. The solid lines represent the corresponding fits according to Eq.~\eqref{phi}. The fit coefficients are listed in Tab.~\ref{Tab1} for sample \#A and in Tab.~\ref{Tab2} for sample \#B in the Appendix~\ref{appendixA}. The inset illustrates the measurement direction of the photocurrent, $J^x$ and $J^y$ with respect to the Hall bar geometry.
	}
	\label{fig2}
\end{figure*}

	\begin{figure}[t] 
	\centering
	\includegraphics[width=\linewidth]{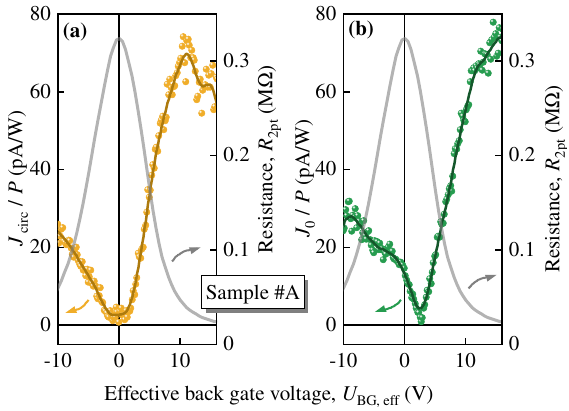}
	\caption{Back gate voltage dependencies of the circular $J_{\rm circ}$, panel (a), and polarization independent $J_0$, panel (b), photocurrent contributions calculated from the measured projections of the individual photocurrents in $x$- and $y$-directions as $J_{\rm circ} = \sqrt{(J^x_{\rm circ})^2+(J^y_{\rm circ})^2}$ and $J_{\rm 0} = \sqrt{(J^x_{\rm 0})^2+(J^y_{\rm 0})^2}$. The correspondingly colored but darker solid traces following their evolution serve as guide for the eyes. The gray trace represents the corresponding two-point resistance (right axes).
	}
	\label{fig3}
\end{figure}

\section{Device fabrication and experimental technique}

The 2D Te flakes were synthesized by the hydrothermal growth method~\cite{Qiu2020}. First, 0.5~g of polyvinylpyrrolidone (PVP) (Sigma-Aldrich) and 0.09~g of Na$_2$TeO$_3$ (Sigma-Aldrich) were dissolved in 32~ml double-distilled water. Then, 3.33~ml of aqueous ammonia solution (25-28\%, weight by weight \%) and 1.67~ml of hydrazine hydrate(80\%,  weight by weight \%) were added under magnetic stirring to form a homogeneous solution. The mixture was sealed in a 50 ml Teflon lined stainless steel autoclave and heated at 180$^\circ$C for 20 hours, then allowed to cool naturally to room temperature. The synthesized 2D Te flakes with a thickness of $\approx 20$~nm were rinsed twice in water before being transferred to a 90~nm SiO$_2$/Si substrate, which served as a back gate, using the Langmuir-Blodgett method. Electron beam lithography was then used to fabricate six-terminal Hall-bar structures, and electron beam evaporation was used to deposit the metal contacts consisting of 20/80~nm Ti/Au. To cap the Te flakes, a 20~nm thick layer of Al$_2$O$_3$ was deposited on top by atomic layer deposition using (CH$_3$)$_3$Al (TMA) and H$_2$O as precursors at 200$^\circ$C. Figure~\ref{fig1} shows the typical cross-section of a Te device [panel (a)] and the top view of sample~\#A [panel (b)]. The micrograph of the second sample under investigation (sample \#B) is shown in Fig.~\ref{figB1} in the Appendix~\ref{appendixA}. The Hall bar is oriented along the $c-$axis of the tellurium, while the $a$-axis is perpendicular to it. In the following, they correspond to the $y-$ and $x-$directions, respectively.

	\begin{figure*}[t]
	\centering
	\includegraphics[width=0.9\linewidth]{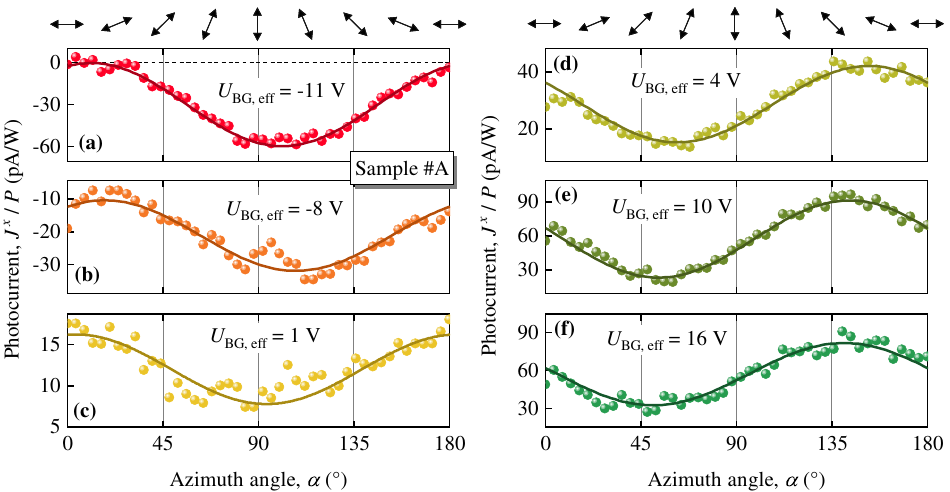}
	\caption{Dependencies of the photocurrent $J^x$ (colored circles) on the azimuthal angle $\alpha$, which defines the orientation of the electric field vector $\bm E$ with respect to the $x-$direction.  The data are shown for several values of the effective back gate voltages ranging from hole ($U_{\rm BG, eff} < 0$) to electron conductivity ($U_{\rm BG, eff} > 0$). The orientations of $\bm E$ of the linearly polarized radiation for several angles $\alpha$ are shown at the top of the figure. The solid lines represent the corresponding fits according to Eq.~\eqref{alpha}. The fit coefficients are given in Tab.~\ref{Tab1} in the Appendix~\ref{appendixA}.
	}
	\label{fig4}
\end{figure*}

 The tellurene samples were investigated at room temperature ($T \approx 300$~K). First the samples were characterized by standard electron transport as a function of the applied back gate voltage. The four-terminal resistance lies in the range of 80 to 100~k$\Omega$ at the charge neutrality point (CNP), while the two-point resistance is about twice as high. For the investigated devices, the CNP was found at high negative gate voltages ($U_{\rm BG}\approx -10$~V). In the following considerations, the applied back gate voltage is presented as an effective voltage with $U_{\rm BG, eff} = U_{\rm BG} - U_{\rm CNP}$, where $U_{\rm CNP}$ is the voltage at which the CNP appears. Consequently, for positive and negative $U_{\rm BG, eff}$ we have electron and hole type conductivity, respectively~\cite{Niu2023b}.  The carrier concentration at room temperature and an effective gate voltage of 20 V is about $1 \times 10^{13}$~cm$^{-2}$, and the mobility $\mu$ is about $600$~cm$^2$/Vs for both electrons and holes.

To excite photocurrents, monochromatic radiation with a frequency $f = 2.54$~THz ($\lambda = 118$~$\mu$m and $E_{\rm ph} = hf = 10.5$~meV) was generated by an optically pumped continuous wave ($cw$) molecular gas laser. The radiation was focused by off-axis parabolic mirrors and applied to the sample under normal incidence to the sample while controlling the polarization state, i.e., the direction and ellipticity of the electric field vector $\textbf{\textit{E}}_\omega$, with $\lambda/2$ or $\lambda/4$ plates. The initial state of polarization ($\alpha = 0, \varphi = 0$) points across the Hall bar, i.e., along the $x$-axis, see the black dashed line in Fig.~\ref{fig1}(c). The laser beam profile at the sample position, measured with a pyroelectric camera\,\cite{Ganichev1999,Herrmann2016}, had a Gaussian shape with a full width at half maximum of $d = 0.9$~mm and reaches intensities up to $I \approx 6.5$~W\,cm$^{-2}$. The area of the laser spot was much larger than the areas of the flakes, confirming uniform irradiation.

The induced photovoltage drop was measured along the $a$ and $c$-axes obtained via the contact pairs 2-6 ($U^x$) and 1-4 ($U^y$), respectively, see Fig.~\ref{fig1}(b). The signals were measured using standard lock-in techniques locked to the chopper frequency, which modulates the $cw$ THz radiation with $f_{\rm chop} = 130$~Hz. The corresponding photocurrents $J$ are related to the photovoltages $U$ according to $J = U/R_{\rm S}$ where $R_{\rm S}(U_{\rm BG, eff})$ is the sample resistance in the 2-6 or 1-4 directions, for $U^x$ and $U^y$, respectively.

\section{Experimental Results}

Photocurrents induced by circularly polarized radiation were observed over the entire gate voltage range investigated in this work. Figure~\ref{fig2} shows the dependence of the photocurrent in samples \#A and \#B measured along ($J^y$) and perpendicular ($J^x$) to the Hall bar. The photocurrent in response to left- and right-handed circularly polarized radiation has different magnitudes, and for some back gate voltages even changes its direction, see, e.g., Fig.~\ref{fig2}(a), (e) and (f).
Its overall polarization dependence is described by 
%
\begin{equation}
	J^{x,y} = J^{x,y}_{\rm circ}\sin(2\varphi) + J^{x,y}_0 + J^{x,y}_{\rm L1} \frac{\cos(4\varphi)+1}{2}+ J^{x,y}_{\rm L2} \frac{\sin(4\varphi)}{2}\,,
	\label{phi}
\end{equation}
where  $J^{x,y}_{\rm circ}$ is the amplitude of the circular photocurrent, $J^{x,y}_0$ is that of the polarization independent contribution, and $J^{x,y}_{\rm L1}$, $J^{x,y}_{\rm L2}$ are the amplitudes of the photocurrent in response to the linearly polarized radiation. Note that the polarization dependencies of the two last components are defined by two Stokes parameters representing the degrees of the linear polarization within the $x, y$ coordinate system and the one rotated by 45$^\circ$ to it. While the overall behavior remains similar as the gate voltage is increased, the individual contributions change, see Fig.~\ref{fig2}. Over the full range of available laser powers, both photocurrent contributions scale linearly with the increase in radiation power $P\propto E_0^2$, where $E_0$ is the electric field of the radiation.

As we discuss below, the photocurrent contributions are caused by nonlinear transport phenomena proportional to the square of the electric field of the radiation and defined by the symmetry of the system. The latter is the lowest for tellurene, which has no nontrivial symmetry elements. For this reason, the direction of the induced photocurrent is arbitrary. Therefore, for the analysis of its functional behavior, in our case the dependencies on the back gate voltage, it is convenient to plot the amplitudes of the photocurrents calculated from their projections in the two perpendicular directions along $x$ and $y$. For example, the magnitude for the circular photocurrent is defined as $J_{\rm circ} = \sqrt{(J^x_{\rm circ})^2+(J^y_{\rm circ})^2}$. The gate voltage dependencies of the circular and the polarization independent photocurrent are shown in Fig.~\ref{fig3} together with the sample resistance. Both photocurrent contributions behave similarly: starting from the hole conductivity region ($U_{\rm BG, eff} < 0$) the photocurrent magnitude decreases, approaches zero and increases significantly in the region of electrons ($U_{\rm BG, eff} > 0$). The gate voltage dependencies of the photocurrent projections are illustrated in Fig.~\ref{figA2} in the Appendix~\ref{appendixA}. 

\begin{figure}[t] 
	\centering
	\includegraphics[width=\linewidth]{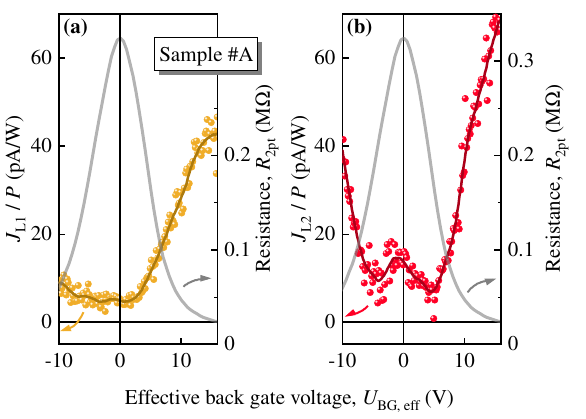}		
	\caption{Back gate voltage dependencies of the linear photocurrent contributions, $J_{\rm L1}$ [panel (a)] and $J_{\rm L2}$ [panel (b)]. Their magnitude was calculated from  the projections of the individual photocurrents measured in the $x$- and $y-$directions using $J_{\rm L1, L2} = \sqrt{(J^x_{\rm L1, L2})^2+(J^y_{\rm L1, L2})^2}$. The correspondingly colored but darker solid traces following their evolution serve as guide for the eyes. The gray traces represent the corresponding two-point resistance (right axes).
	}
	\label{fig5}
\end{figure}

While the photocurrent induced by linearly polarized radiation contributes significantly to the polarization dependencies, obtained by rotating the $\lambda$/4 wave plate, they can be easily studied using linearly polarized radiation. Corresponding dependencies are shown exemplary in $x$-direction in Fig.~\ref{fig4}. In this experimental setup the polarization dependencies are defined as
\begin{equation}
	J^{x,y} = J^{x,y}_0 + J^{x,y}_{\rm L1}\cos(2\alpha) + J^{x,y}_{\rm L2}\sin(2\alpha)\,,
	\label{alpha}
\end{equation} 
with the same amplitudes $J^{x,y}_0$, $J^{x,y}_{\rm L1}$, and $J^{x,y}_{\rm L2}$ as in Eq.~\eqref{phi}. Note that the conversion of Eq.~\eqref{phi} into Eq.~\eqref{alpha} and vice versa is related to that of the Stokes parameters, see, e.g., Refs.~\cite{Saleh2019, Belkov2005}. Figure~\ref{fig4} shows that increasing the back gate voltage changes both the amplitude and the phase of the polarization-dependent photocurrent. This is due to the different gate voltage dependencies of $J_{\rm L1}$ and $J_{\rm L2}$. The same behavior was observed along the $y$-direction, see the Fig.~\ref{figA3} in the Appendix~\ref{appendixA}. Similar to the photocurrents $J_{\rm circ}$ and $J_{\rm 0}$ we calculate the individual gate voltage dependencies of $J_{\rm L1}$ and $J_{\rm L2}$ according to $J_{\rm L1, L2} = \sqrt{(J^x_{\rm L1, L2})^2+(J^y_{\rm L1, L2})^2}$. Figure~\ref{fig5} shows that both contributions indeed depend differently on the gate voltage and that the photocurrent $J_{\rm L2}$ is dominating over the whole range. Note that both photocurrents approach minima near the CNP.

\section{Theory and Discussion}

In bulk Te, the linear and circular currents are expected only in certain crystallographic directions. In $(10\bar{1}0)$ 2D tellurene, the point symmetry group is $C_2$, with the only nontrivial symmetry element being rotation by the angle $\pi$ about the in-plane axis $C_2 \parallel [1\bar{2}10]$ ($a$ axis). The presence of a substrate makes the $\pm z-$directions inequivalent, and the $C_2$ operation does not transform the tellurene sample to itself. As a result, the point symmetry is reduced to $C_1$ where any nontrivial symmetry elements are excluded.
In systems with such a low symmetry excitation with homogeneous radiation at normal incidence results in a dc current, which is described by
\begin{equation}
	\label{phenom}
	j^{x, y} = ( \gamma^{x, y} P_{\text{circ}} + A^{x, y} +C^{x, y} P_{\text{L1}}  + S^{x, y} P_{\text{L2}}) E_0^2\,. 
\end{equation}
Here $j^{x, y} \propto J^{x, y}$ is the photocurrent density projection on the axis $x$ or $y$
in the structure plane, the radiation's electric field is $\bm E_\omega = \bm e E_0\exp(-i\omega t) + \text{c.c.}$ where $\bm e$ is the polarization unit vector and $E_0$ is the amplitude, $P_{\text{circ}}=-2{\rm Im}(e_xe_y^*)$, $P_{\text{L1}}=|e_x|^2-|e_y|^2$, $P_{\text{L2}}=2{\rm Re}(e_xe_y^*)$  are the Stokes parameters of the radiation~\cite{Saleh2019},  and $\gamma^{x,y}$, $A^{x,y}$, $C^{x,y}$, and $S^{x,y}$  are eight coefficients that are linearly independent due to the absence of any nontrivial symmetry operation. Indeed, our experimental results show that all eight coefficients have comparable magnitudes, resulting in photocurrents in $a$- and $c$-directions, see Tabs.~\ref{Tab1} and \ref{Tab2} in the Appendix~\ref{appendixA}.
Therefore, the experiments clearly demonstrate that we are dealing with a 2D system, which agrees with previous results on quantum Hall effect measurements on the similar structures~\cite{Qiu2020,Niu2021}.

Now we discuss the observed polarization dependencies. The introduced parameters $P_{\text{circ}}$, $P_{\text{L1}}$, and $P_{\text{L2}}$ describe the polarization state, which varies by the rotation of the wave plates in our experiments. These parameters are given by the trigonometric functions in Eqs.~\eqref{phi} and \eqref{alpha} describing the polarization dependencies of the photocurrent contributions associated with $J_{\rm circ}^{x,y}$, $J_{\rm L1}^{x,y}$ and $J_{\rm L2}^{x,y}$~\cite{Belkov2005}. Consequently, the photocurrent described by the phenomenological Eq.~\eqref{phenom} is in full agreement with the experimental results showing the above polarization dependencies presented in Figs.~\ref{fig2} and~\ref{fig4}, as well as in Figs.~\ref{figA3}, \ref{figB2}, and \ref{figB3} in the Appendix~\ref{appendixA}. Particularly interesting is the analysis of the polarization dependencies shown in Fig.~\ref{fig4}, because it allows one to clearly separate the high-frequency NLH and NLL contributions. Indeed, $J^x$ measured at $\alpha = 90^\circ$ $(\bm E \parallel y)$ gives the NLH current perpendicular to the radiation's electric field, while $J^x$ measured at $\alpha = 0$ $(\bm E \parallel x)$ corresponds to the NLL.  Figure~\ref{fig4} shows that they are comparable, which is not surprising, see also Tabs.~\ref{Tab1}, \ref{Tab2} and Figs.~\ref{figA3}, \ref{figB3} in the Appendix~\ref{appendixA}. Note that for $J^y$, NLH and NLL are obtained for  $\alpha = 0$ and $\alpha = 90^\circ$, respectively.

The investigated photocurrents are excited  by THz radiation with a photon energy of a few meV, which is lower than the energy gap of tellurene. Consequently, they are caused by the free carrier Drude-like radiation absorption. At room temperature in tellurene the THz radiation excites only one carrier type, either holes or electrons, depending on the gate voltage~\cite{Qiu2020}. 
Under these conditions, the THz-induced  photocurrents  given by Eq.~\eqref{phenom} are 
described by the Boltzmann kinetic equation for the electron distribution function $f_{\bm k}$. Here, $\bm k$ is the two-dimensional electron wavevector and the equation has the following form
\begin{equation}
\label{kin_eq}
{\partial f_{\bm k}\over \partial t} + {q\over \hbar}\bm E_\omega \cdot {\partial f_{\bm k}\over \partial \bm k} = \sum_{\bm k'} (W_{\bm k \bm k'}f_{\bm k'}-W_{\bm k' \bm k}f_{\bm k})\,,
\end{equation}
where $W_{\bm k' \bm k}$ is the probability of scattering between the electron states with wavevectors $\bm k$ and $\bm k'$ and $q$ is the carrier charge being $\pm|e|$ for holes and electrons, respectively.
The electric current density is calculated as
\begin{equation}
\label{j_def}
\bm j = g_sg_v q\sum_{\bm k}f_{\bm k} \bm v_{\bm k}\,,
\end{equation}
where $g_s$ and $g_v$ are the spin and valley degeneracies,
and $\bm v_{\bm k}$ is the electron velocity. The distribution function  $f_{\bm k}$ is quadratic in $q\bm E_\omega$, corresponding to the photocurrent's linear dependence on the radiation power observed in the experiment. Equation~\eqref{j_def} also shows that the photocurrent is proportional to  the third power of the carrier charge. Consequently, it changes its direction  when passing through the CNP and approaches zero value at $U_{\rm BG, eff} = 0$. The latter is evident from the experimental data, see Figs.~\ref{fig3} and \ref{figA2} in the Appendix~\ref{appendixA}. 

In the semiclassical approach, there are three microscopic mechanisms of THz radiation-induced photocurrents. As mentioned in the introduction, these are the intrinsic Berry curvature dipole (BCD) mechanism and extrinsic contributions  due to side jumps of electron wave packets occurring in momentum scattering and skew scattering processes~\cite{Golub2020,Ortix2021}. 
The BCD mechanism of the photocurrent is accounted for in the anomalous velocity linear in the electric field of the radiation
\begin{equation}
\label{v_anom}
\bm v^{\rm anom}_{\bm k}=\bm \Omega_{\bm k} \times \bm E_\omega\,.
\end{equation}
Here, the Berry curvature is $\bm \Omega_{\bm k} = \bm \nabla_{\bm k}\times i\braket{u_{\bm k}|{\nabla_{\bm k} u_{\bm k}}}$, where $u_{\bm k}$ is the electron Bloch amplitude  for the energy band in which the radiation is absorbed.
Finding the usual correction to the distribution function $f_1 \propto E_\omega$ from the Boltzmann Eq.~\eqref{kin_eq} with an ordinary scattering probability $W^0_{\bm k' \bm k}=W^0_{\bm k \bm k'}$, the BCD contribution to the photocurrent is then calculated by Eq.~\eqref{j_def} with $\bm v_{\bm k}=\bm v^{\rm anom}_{\bm k}$.
For elastic scattering, $W^0_{\bm k' \bm k}=(2\pi/\hbar)\mathcal K(|\bm k'-\bm k|)\delta(\varepsilon_{\bm k}-\varepsilon_{\bm k'})$, where $\varepsilon_{\bm k}$ is the energy dispersion in the band, and $\mathcal K(q)$ is the disorder potential correlator. The considered mechanism is responsible for all detected currents: circular, linear, and polarization independent. Note that, as can be seen from Eq.~\eqref{v_anom}, the BCD mechanism at linear polarization contributes only to the perpendicular electric current, i.e., to the NLH.
We also find that the circular photocurrent is present at frequency $\omega$ comparable or higher than the transport relaxation rate. 

Another photocurrent  mechanism is the side-jump caused by the electron wavepacket shifts $\bm r_{\bm k' \bm k}$ that occur during scattering and is given by
\begin{equation}
\bm r_{\bm k' \bm k} = -(\bm \nabla_{\bm k'}+\bm \nabla_{\bm k})\Phi_{\bm k' \bm k} + \bm A_{\bm k'} - \bm A_{\bm k}\,,
\end{equation}
where $\Phi_{\bm k' \bm k}$ is the phase of the matrix element of the scattering $\bm k \to \bm k'$, and the Berry connection $\bm A_{\bm k}= i\braket{u_{\bm k}|{\nabla_{\bm k} u_{\bm k}}}$.

There are two contributions to the photocurrent due to side jumps. One comes from the side-jump accumulation, which leads to the velocity correction
\begin{equation}
\bm v^{\rm sj}_{\bm k}=\sum_{\bm k'}W^0_{\bm k' \bm k}\bm r_{\bm k' \bm k}\,.
\end{equation}
Then by finding the correction to the distribution function $f_2 \propto E^2_\omega$ from the Boltzmann Eq.~\eqref{kin_eq} with 
the scattering probability $W^0_{\bm k' \bm k}$, the side-jump accumulation contribution to the photocurrent is calculated by Eq.~\eqref{j_def} with $\bm v_{\bm k}=\bm v^{\rm sj}_{\bm k}$. This contribution is insensitive to the radiation helicity and contributes to the constants $A^{x, y}$, $C^{x, y}$ and $S^{x, y}$ in Eq.~\eqref{phenom}, i.e., to the polarization-independent and linear polarization dependent photocurrents.

Another correction to the distribution function comes from 
the linear in the electric field scattering probability $W^{\rm sj}_{\bm k' \bm k}$
caused by the correction to the energy conservation law due to the work of the field at the side jump:
\begin{equation}
W^{\rm sj}_{\bm k' \bm k}={2\pi\over\hbar}\mathcal K(|\bm k'-\bm k|)(q\bm E_\omega \cdot \bm r_{\bm k' \bm k}) \partial_{\varepsilon_{\bm k}}\delta(\varepsilon_{\bm k}-\varepsilon_{\bm k'})\,.
\end{equation}
Iterating the kinetic Eq.~\eqref{kin_eq} with $W_{\bm k' \bm k}=W^0_{\bm k' \bm k}+W^{\rm sj}_{\bm k' \bm k}$ in $E_\omega$, one finds the so-called anomalous distribution $f^{\rm adist} \propto E_\omega^2$. Then the corresponding contribution to the photocurrent is calculated by Eq.~\eqref{j_def} with $f_{\bm k}=f^{\rm adist}$ and an ordinary velocity $\bm v_{\bm k}=\hbar^{-1}\bm \nabla_{\bm k}\varepsilon_{\bm k}$. This mechanism also gives rise to all observed photocurrent contributions including the helicity dependent photocurrent.

Finally, the linear photocurrent can also be caused by skew scattering, previously considered for dc NLH and LPGE~\cite{Du2021,Otteneder2020}. In the case of the circular photocurrent, BCD and side-jump contributions dominate for THz frequencies and room temperature, see Appendix~\ref{Skew}.

To sum up, the obtained expressions describe nonlinear transport in 2D tellurene. As mentioned above, they yield circular~\footnote{Note that the circular photocurrent (CPGE) excited by infrared radiation with photon energy exceeding the energy gap and caused by direct interband optical transitions has been observed in Ref.~\cite{Niu2023a}.} and polarization-independent photocurrents (see Figs.~\ref{fig2}, \ref{fig3}, \ref{figA2}, \ref{figB2}, and \ref{figB4}), as well as linear photocurrents (see Figs.~\ref{fig4}, \ref{fig5}, \ref{figA3}, and \ref{figB3} in the Appendix~\ref{appendixA}). While the BCD mechanism at linear polarization gives the nonlinear current perpendicular to the THz electric field, the side-jump mechanism contributes to all photocurrents.
For circular polarization, both BCD and side-jump mechanisms contribute to the nonlinear currents~\footnote{Note that for bulk tellurium the BCD contribution to the circular photocurrent along $C_3$-axis was calculated in Ref.~\cite{Tsirkin2018}.}.

\section{Conclusion}

We have studied the direct current excited by polarized terahertz radiation in tellurene structures. The current, which is quadratic in the electric field, belongs to the class of nonlinear electron transport effects. Photocurrents sensitive to radiation helicity and orientation of the electric field vector as well as gate voltage were observed. We attribute the THz-induced currents to two mechanisms arising from the low spatial symmetry of 2D tellurene: the intrinsic Berry curvature dipole and extrinsic side-jumps during momentum scattering. Our results show that tellurene structures reduce to the lowest symmetry group and that the perpendicular photocurrent, similar to the nonlinear Hall effect at high frequencies, is free of linear electric field contributions. This allows free access to the longitudinal current and polarization dependent effects at THz frequencies. Beyond that, the use of  THz frequencies facilitates the observation of photocurrents, which are sensitive to the direction of rotation of the ac electric field.

\section{Acknowledgments}

The financial support of the Deutsche Forschungsgemeinschaft (DFG, German Research Foundation) via Project-ID 521083032 (Ga501/19), the Volkswagen Stiftung Program (97738) is gratefully acknowledged.  
S.D.G. is also grateful for the support of the European Union (ERC-ADVANCED “TERAPLASM,” Project No. 101053716). J.G.-S. acknowledges the support from DFG GU 2528/1-1 695298.

\appendix

\section{Additional data}
\label{appendixA}

\begin{figure}[t] 
	\centering
	\includegraphics[width=\linewidth]{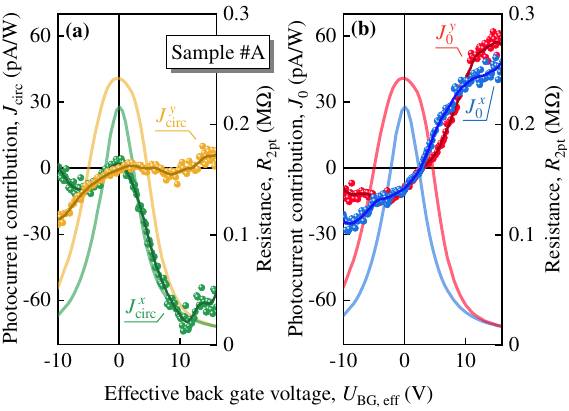}
	\caption{Back gate voltage dependencies of the circular ($J_{\rm circ}$) and polarization independent ($J_0$) photocurrents measured across ($J^x_{\rm circ}$, along the $a-$direction) and along ($J^y_{\rm circ}$, along the $c-$direction) the Hall bar. The solid lines following their evolution serve as guide for the eyes. The two-terminal resistances are shown by correspondingly colored curves (right axes).
	}
	\label{figA2}
\end{figure}

Figure~\ref{figA2} shows the gate voltage dependence of the circular and polarization independent photocurrent projections measured in $x-$ and $y-$ directions.

\begin{figure}[t] 
	\centering
	\includegraphics[width=\linewidth]{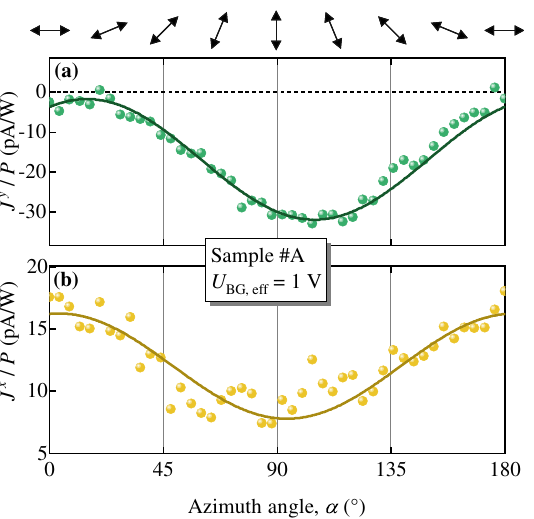}
	\caption{Photocurrents  $J^y$, panel (a), and  $J^x$, panel (b),  as a function of the azimuth angle $\alpha$ defining the orientation of the electric field of linearly polarized radiation. The data are shown for $U_{\rm BG, eff} = 1$~V. The circles show the measured traces, whereas the solid lines show the corresponding fits after Eq.~\ref{alpha} of the main text. The fit coefficients are given in Tab.~\ref{Tab1}. The symbols on top illustrate  polarization states for several azimuth angles $\alpha$.}
	\label{figA3}
\end{figure}

Figure~\ref{figA3} shows the  dependence of the photocurrent on the orientation of the radiation electric field vector in respect to the $y-$axis measured in $x-$ and $y-$ directions.

\begin{figure}[t]
	\centering
	\includegraphics[width=\linewidth]{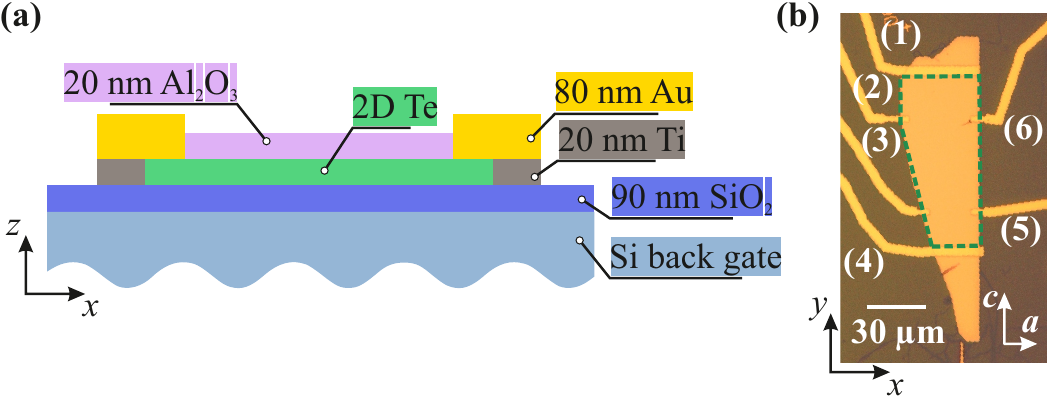}
	\caption{Panel (a) shows a typical cross section of the 2D Te devices under investigation. Panel (b) shows an optical microphotograph of sample \#B and the corresponding contact numbering. The green dashed pentagon highlights the Hall bar.}
	\label{figB1}
\end{figure}

\begin{figure}[t] 
	\centering
	\includegraphics[width=\linewidth]{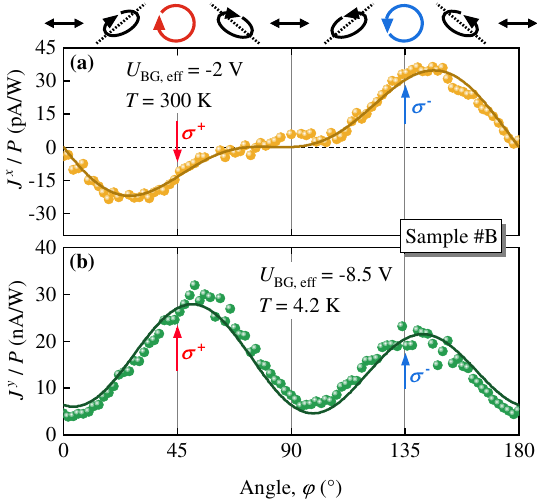}
	\caption{Photocurrents $J^x$, panel (a), and  $J^y$, panel (b), measured in sample \#B as a function of the angle $\varphi$ defining the degree of circular polarization. The former was measured at $T = 300$~K, and the latter was obtained at $T = 4.2$~K. The circles represent the measured traces, whereas the solid lines show the corresponding fits after Eq.~\ref{phi} of the main text. The fit coefficients are given in Tab.~\ref{Tab2}. The symbols on top illustrate several polarization states at several angles $\varphi$, and blue arrows label the right-handed ($\sigma^+$) and left-handed ($\sigma^-$) circular polarized radiation.}
	\label{figB2}
\end{figure}

\begin{figure}[t] 
	\centering
	\includegraphics[width=\linewidth]{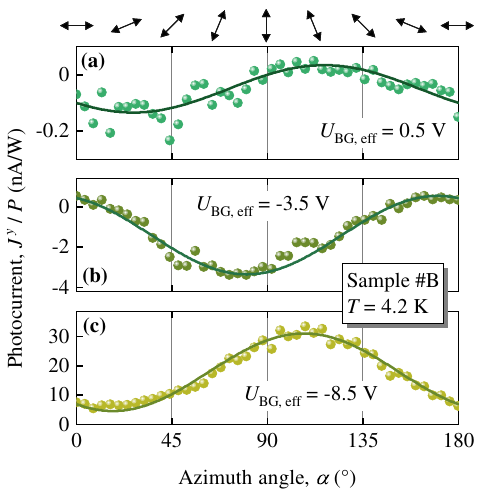}
	\caption{Azimuth angle dependencies of the photocurrent $J^y$ (colored circles) obtained for several examples of effective back gate voltages. Selected polarization states are shown on top of the figure. The solid lines represent the corresponding fits after Eq.~\ref{alpha} of the main text. The fit coefficients are given in Tab.~\ref{Tab2}.}
	\label{figB3}
\end{figure}

\begin{figure}[t] 
	\centering
	\includegraphics[width=\linewidth]{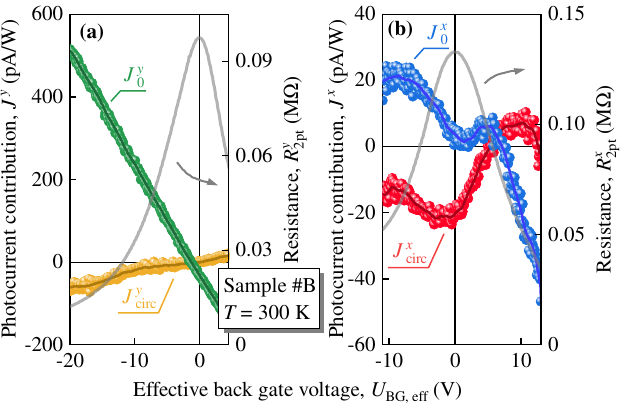}
	\caption{Back gate voltage dependencies of the circular $J_{\rm circ}$ and the polarization independent $J_0$ photocurrent contributions along the $y$, panel (a), and the $x$, panel (b), directions. The solid lines following their evolution serve as guide for the eyes. The two-terminal resistances are shown by the grey shaded curves (right axes).}
	\label{figB4}
\end{figure}

Similar data were obtained for the second sample (sample \#B). Figure~\ref{figB1} shows the cross-section [panel (a)] and a micrograph [panel (b)] of sample \#B. Its parameters and layer thicknesses are very similar to that of sample \#A. Figures~\ref{figB2} - \ref{figB4} present the dependence of the photocurrents on the radiation helicity, the polarization orientation and the back gate voltage, respectively in both $x$- and $y$-directions.

All polarization dependencies on the azimuth angle $\alpha$ and $\varphi$ were fitted after Eqs.~\ref{phi} and \ref{alpha} of the main text. The fit coefficients used for samples \#A and \#B are listed in Tables~\ref{Tab1} and \ref{Tab2}, respectively. For convenience the values are related to the figure number and the corresponding back gate voltage.

\begin{table*}
	\centering
	\begin{tabularx}{\textwidth}{XXXXXX}
		\toprule[0.05cm]\addlinespace[0.2cm]
		Figure No. & $U_{\rm BG, eff}$ (V) & $J_0$ (pA/W) & $J_{\rm L1}$ (pA/W) & $J_{\rm L2}$ (pA/W) & $J_{\rm circ}$ (pA/W) 					\\
		\midrule[0.025cm]\addlinespace[0.1cm]
		Figure~\ref{fig2}(a) & -11  & -7.7   & 24.3     & 2.0       &-18.7         \\
		Figure~\ref{fig2}(b) & -5                & -8.6   & 6.7           & 2.0       & -3.6         \\
		Figure~\ref{fig2}(c) & 16                & 51.4   & -51.2         & -19.1     & 12.3         \\
		Figure~\ref{fig2}(d) & 10                & 31.9   & -19.7         & 5.3	     & -11.8        \\
		Figure~\ref{fig2}(e) & 10   				&	17.1 & -23.1    	 & 10.9      & -35.4    \\	
		\midrule[0.025cm]\addlinespace[0.1cm]
		Figure~\ref{fig4}(a) & -11  & -29.9  & 27.4     & 11.3      & --           \\
		Figure~\ref{fig4}(b) & -8                & -21.1  & 8.8           & 6.3       & --           \\
		Figure~\ref{fig4}(c) & 1                 & 12.0   & 4.2           & 0.6       & --  			\\
		Figure~\ref{fig4}(d) & 4                 & 28.8   & 7.6           & -11.1     & --   	    \\
		Figure~\ref{fig4}(e) & 10                & 57.4   & 9.8           & -32.7     & --   	    \\
		Figure~\ref{fig4}(f) & 16                & 57.2   & 4.9           & -24.1     & --           \\
		\midrule[0.025cm]\addlinespace[0.1cm]
		Figure~\ref{figA3}(a)& 1    &	-16.8 & 13.1     & 7.4       & --       \\	
		Figure~\ref{figA3}(b)& 1    & 12.0    & 4.2      & 0.6       & --       \\
		\midrule[0.025cm]\addlinespace[0.1cm]
		\addlinespace[0.1cm]\bottomrule[0.05cm]
	\end{tabularx}
	\caption{Coefficients of the fits after Eqs.~\ref{phi} and \ref{alpha} of the main text for sample \#A. The first two columns give the corresponding figure numbers (of the main text and Appendix) and the effective back gate voltages of the fitted curves.\vspace{1cm}}
	\label{Tab1}
\end{table*}

\begin{table*}
	\centering
	\begin{tabularx}{\textwidth}{XXXXXX}
		\toprule[0.05cm]\addlinespace[0.2cm]
		Figure No. & $U_{\rm BG, eff}$ (V) & $J_0$ (pA/W) & $J_{\rm L1}$ (pA/W) & $J_{\rm L2}$ (pA/W) & $J_{\rm circ}$ (pA/W) 					\\
		\midrule[0.025cm]\addlinespace[0.1cm]
		Figures~\ref{fig2}(f) \& \ref{figB2}(a)& -2    &	8.6 & -8.4     & -20.6       & -22.0    \\	
		Figure~\ref{figB2}(b)& -8.5  & 2.4$\times10^4$ & -1.7$\times10^4$    & -8.2 $\times10^3$       & 3.3$\times10^3$      \\
		\midrule[0.025cm]\addlinespace[0.1cm]
		Figure~\ref{figB3}(a)& 0.5   &	-49.6 & -50.0     & -67.6       & --    \\	
		Figure~\ref{figB3}(b)& -3.5  & -2.6$\times10^2$ & 4.7$\times10^2$    & -1.2 $\times10^3$       & --      \\
		Figure~\ref{figB3}(c)& -8.5  & 1.8$\times10^4$ & -1.1$\times10^4$    & -7.5 $\times10^3$       & --      \\
		\addlinespace[0.1cm]\bottomrule[0.05cm]
	\end{tabularx}
	\caption{Coefficients of the fits after Eqs.~\ref{phi} and \ref{alpha} of the main text for sample \#B. The first two columns give the corresponding figure numbers (of the main text and Appendix) and the effective back gate voltages of the fitted curves.}
	\label{Tab2}
\end{table*}

\section{Skew scattering mechanism}
\label{Skew}

The skew-scattering is caused by the fact that, in  
noncentrosymmetric systems, the scattering probability does not satisfy the principle of detailed balance: the probability of $\bm k' \rightarrow \bm k$ process, $W_{\bm k \bm k'}$ differs from that of inverse process, $W_{\bm k' \bm k}$.
Therefore it is possible to make a splitting to the symmetric and asymmetric parts: $W_{\bm k \bm k'}= W_{\bm k \bm k'}^0 + W_{\bm k \bm k'}^{\rm sk}$, where $W_{\bm k \bm k'}^{0, {\rm sk}} = \pm W_{\bm k' \bm k}^{0, {\rm sk}}$~\cite{Otteneder2020}. The skew-scattering induced correction to the distribution function is obtained by two iterations of the kinetic equation in powers of the electric field $\bm E_\omega$ and one iteration in $W_{\bm k \bm k'}^{\rm sk}$. Then the skew-scattering contribution to the photocurrent is calculated by Eq. (7) of the main text with an ordinary velocity. In systems of $C_1$-symmetry it gives rise to all contributions in Eq. (5) of the main text.

At high frequency used in the experiments  we have $\omega \tau > 1$, where
$\tau$ is a  transport relaxation time.
Therefore, the skew-scattering based mechanism of the circular photocurrent is suppressed and yield negligible contribution in respect to the BCD and side-jump driven photocurrents. Indeed, this is valid for both terms in the skew scattering probability. A so-called conventional skew scattering occurs with non-Gaussian disorder and appears in the third order in the scattering potential. It is present at low temperatures for scattering by impurities, but is absent for acoustic phonon scattering when single-phonon processes are dominant~\cite{Glazov2020a,Glazov2020}. It is also valid, for another, the so-called intrinsic skew scattering, occurs at any potential and is proportional to the fourth power of the disorder potential.
Summarizing, at high frequencies, $\omega\tau \gg 1$, the intrinsic skew scattering contribution to circular photocurrent is suppressed as it scales with $(\omega\tau)^{-2}$ of the BCD and side-jump contribution~\cite{Golub2020}.

\bibliography{all_lib1.bib}

\begin{thebibliography}{51}%
\makeatletter
\providecommand \@ifxundefined [1]{%
 \@ifx{#1\undefined}
}%
\providecommand \@ifnum [1]{%
 \ifnum #1\expandafter \@firstoftwo
 \else \expandafter \@secondoftwo
 \fi
}%
\providecommand \@ifx [1]{%
 \ifx #1\expandafter \@firstoftwo
 \else \expandafter \@secondoftwo
 \fi
}%
\providecommand \natexlab [1]{#1}%
\providecommand \enquote  [1]{``#1''}%
\providecommand \bibnamefont  [1]{#1}%
\providecommand \bibfnamefont [1]{#1}%
\providecommand \citenamefont [1]{#1}%
\providecommand \href@noop [0]{\@secondoftwo}%
\providecommand \href [0]{\begingroup \@sanitize@url \@href}%
\providecommand \@href[1]{\@@startlink{#1}\@@href}%
\providecommand \@@href[1]{\endgroup#1\@@endlink}%
\providecommand \@sanitize@url [0]{\catcode `\\12\catcode `\$12\catcode
  `\&12\catcode `\#12\catcode `\^12\catcode `\_12\catcode `\%12\relax}%
\providecommand \@@startlink[1]{}%
\providecommand \@@endlink[0]{}%
\providecommand \url  [0]{\begingroup\@sanitize@url \@url }%
\providecommand \@url [1]{\endgroup\@href {#1}{\urlprefix }}%
\providecommand \urlprefix  [0]{URL }%
\providecommand \Eprint [0]{\href }%
\providecommand \doibase [0]{https://doi.org/}%
\providecommand \selectlanguage [0]{\@gobble}%
\providecommand \bibinfo  [0]{\@secondoftwo}%
\providecommand \bibfield  [0]{\@secondoftwo}%
\providecommand \translation [1]{[#1]}%
\providecommand \BibitemOpen [0]{}%
\providecommand \bibitemStop [0]{}%
\providecommand \bibitemNoStop [0]{.\EOS\space}%
\providecommand \EOS [0]{\spacefactor3000\relax}%
\providecommand \BibitemShut  [1]{\csname bibitem#1\endcsname}%
\let\auto@bib@innerbib\@empty
\bibitem [{\citenamefont {Wu}\ \emph {et~al.}(2018{\natexlab{a}})\citenamefont
  {Wu}, \citenamefont {Qiu}, \citenamefont {Wang}, \citenamefont {Wang},\ and\
  \citenamefont {Ye}}]{Wu2018b}%
  \BibitemOpen
  \bibfield  {author} {\bibinfo {author} {\bibfnamefont {W.}~\bibnamefont
  {Wu}}, \bibinfo {author} {\bibfnamefont {G.}~\bibnamefont {Qiu}}, \bibinfo
  {author} {\bibfnamefont {Y.}~\bibnamefont {Wang}}, \bibinfo {author}
  {\bibfnamefont {R.}~\bibnamefont {Wang}},\ and\ \bibinfo {author}
  {\bibfnamefont {P.}~\bibnamefont {Ye}},\ }\bibfield  {title} {\bibinfo
  {title} {Tellurene: its physical properties, scalable nanomanufacturing, and
  device applications},\ }\href {https://doi.org/10.1039/c8cs00598b} {\bibfield
   {journal} {\bibinfo  {journal} {Chemical Society Reviews}\ }\textbf
  {\bibinfo {volume} {47}},\ \bibinfo {pages} {7203} (\bibinfo {year}
  {2018}{\natexlab{a}})}\BibitemShut {NoStop}%
\bibitem [{\citenamefont {Wu}\ \emph {et~al.}(2018{\natexlab{b}})\citenamefont
  {Wu}, \citenamefont {Huang}, \citenamefont {Wang}, \citenamefont {Zhao},
  \citenamefont {Ma}, \citenamefont {Xiang}, \citenamefont {Li}, \citenamefont
  {Ponraj}, \citenamefont {Dhanabalan},\ and\ \citenamefont {Zhang}}]{Wu2018a}%
  \BibitemOpen
  \bibfield  {author} {\bibinfo {author} {\bibfnamefont {L.}~\bibnamefont
  {Wu}}, \bibinfo {author} {\bibfnamefont {W.}~\bibnamefont {Huang}}, \bibinfo
  {author} {\bibfnamefont {Y.}~\bibnamefont {Wang}}, \bibinfo {author}
  {\bibfnamefont {J.}~\bibnamefont {Zhao}}, \bibinfo {author} {\bibfnamefont
  {D.}~\bibnamefont {Ma}}, \bibinfo {author} {\bibfnamefont {Y.}~\bibnamefont
  {Xiang}}, \bibinfo {author} {\bibfnamefont {J.}~\bibnamefont {Li}}, \bibinfo
  {author} {\bibfnamefont {J.~S.}\ \bibnamefont {Ponraj}}, \bibinfo {author}
  {\bibfnamefont {S.~C.}\ \bibnamefont {Dhanabalan}},\ and\ \bibinfo {author}
  {\bibfnamefont {H.}~\bibnamefont {Zhang}},\ }\bibfield  {title} {\bibinfo
  {title} {{2D} tellurium based high‐performance all‐optical nonlinear
  photonic devices},\ }\href {https://doi.org/10.1002/adfm.201806346}
  {\bibfield  {journal} {\bibinfo  {journal} {Adv. Funct. Mater.}\ }\textbf
  {\bibinfo {volume} {29}},\ \bibinfo {pages} {1806346} (\bibinfo {year}
  {2018}{\natexlab{b}})}\BibitemShut {NoStop}%
\bibitem [{\citenamefont {Shi}\ \emph {et~al.}(2020)\citenamefont {Shi},
  \citenamefont {Cao}, \citenamefont {Khan}, \citenamefont {Tareen},
  \citenamefont {Liu}, \citenamefont {Liang}, \citenamefont {Zhang},
  \citenamefont {Ma}, \citenamefont {Guo}, \citenamefont {Luo},\ and\
  \citenamefont {Zhang}}]{Shi2020}%
  \BibitemOpen
  \bibfield  {author} {\bibinfo {author} {\bibfnamefont {Z.}~\bibnamefont
  {Shi}}, \bibinfo {author} {\bibfnamefont {R.}~\bibnamefont {Cao}}, \bibinfo
  {author} {\bibfnamefont {K.}~\bibnamefont {Khan}}, \bibinfo {author}
  {\bibfnamefont {A.~K.}\ \bibnamefont {Tareen}}, \bibinfo {author}
  {\bibfnamefont {X.}~\bibnamefont {Liu}}, \bibinfo {author} {\bibfnamefont
  {W.}~\bibnamefont {Liang}}, \bibinfo {author} {\bibfnamefont
  {Y.}~\bibnamefont {Zhang}}, \bibinfo {author} {\bibfnamefont
  {C.}~\bibnamefont {Ma}}, \bibinfo {author} {\bibfnamefont {Z.}~\bibnamefont
  {Guo}}, \bibinfo {author} {\bibfnamefont {X.}~\bibnamefont {Luo}},\ and\
  \bibinfo {author} {\bibfnamefont {H.}~\bibnamefont {Zhang}},\ }\bibfield
  {title} {\bibinfo {title} {Two-dimensional tellurium: Progress, challenges,
  and prospects},\ }\href {https://doi.org/10.1007/s40820-020-00427-z}
  {\bibfield  {journal} {\bibinfo  {journal} {Nano-Micro Lett.}\ }\textbf
  {\bibinfo {volume} {12}},\ \bibinfo {pages} {99} (\bibinfo {year}
  {2020})}\BibitemShut {NoStop}%
\bibitem [{\citenamefont {Qiu}\ \emph {et~al.}(2022)\citenamefont {Qiu},
  \citenamefont {Charnas}, \citenamefont {Niu}, \citenamefont {Wang},
  \citenamefont {Wu},\ and\ \citenamefont {Ye}}]{Qiu2022}%
  \BibitemOpen
  \bibfield  {author} {\bibinfo {author} {\bibfnamefont {G.}~\bibnamefont
  {Qiu}}, \bibinfo {author} {\bibfnamefont {A.}~\bibnamefont {Charnas}},
  \bibinfo {author} {\bibfnamefont {C.}~\bibnamefont {Niu}}, \bibinfo {author}
  {\bibfnamefont {Y.}~\bibnamefont {Wang}}, \bibinfo {author} {\bibfnamefont
  {W.}~\bibnamefont {Wu}},\ and\ \bibinfo {author} {\bibfnamefont {P.~D.}\
  \bibnamefont {Ye}},\ }\bibfield  {title} {\bibinfo {title} {The resurrection
  of tellurium as an elemental two-dimensional semiconductor},\ }\href
  {https://doi.org/10.1038/s41699-022-00293-w} {\bibfield  {journal} {\bibinfo
  {journal} {npj 2D Mater. Appl.}\ }\textbf {\bibinfo {volume} {6}},\ \bibinfo
  {pages} {17} (\bibinfo {year} {2022})}\BibitemShut {NoStop}%
\bibitem [{\citenamefont {Zha}\ \emph {et~al.}(2024)\citenamefont {Zha},
  \citenamefont {Dong}, \citenamefont {Huang}, \citenamefont {Xia},
  \citenamefont {Tong}, \citenamefont {Liu}, \citenamefont {Chan},
  \citenamefont {Ho}, \citenamefont {Zhao}, \citenamefont {Chai},\ and\
  \citenamefont {Tan}}]{Zha2024}%
  \BibitemOpen
  \bibfield  {author} {\bibinfo {author} {\bibfnamefont {J.}~\bibnamefont
  {Zha}}, \bibinfo {author} {\bibfnamefont {D.}~\bibnamefont {Dong}}, \bibinfo
  {author} {\bibfnamefont {H.}~\bibnamefont {Huang}}, \bibinfo {author}
  {\bibfnamefont {Y.}~\bibnamefont {Xia}}, \bibinfo {author} {\bibfnamefont
  {J.}~\bibnamefont {Tong}}, \bibinfo {author} {\bibfnamefont {H.}~\bibnamefont
  {Liu}}, \bibinfo {author} {\bibfnamefont {H.~P.}\ \bibnamefont {Chan}},
  \bibinfo {author} {\bibfnamefont {J.~C.}\ \bibnamefont {Ho}}, \bibinfo
  {author} {\bibfnamefont {C.}~\bibnamefont {Zhao}}, \bibinfo {author}
  {\bibfnamefont {Y.}~\bibnamefont {Chai}},\ and\ \bibinfo {author}
  {\bibfnamefont {C.}~\bibnamefont {Tan}},\ }\bibfield  {title} {\bibinfo
  {title} {Electronics and optoelectronics based on tellurium},\ }\href
  {https://doi.org/10.1002/adma.202408969} {\bibfield  {journal} {\bibinfo
  {journal} {Adv. Mater.}\ ,\ \bibinfo {pages} {2408969}} (\bibinfo {year}
  {2024})}\BibitemShut {NoStop}%
\bibitem [{\citenamefont {Wu}\ \emph {et~al.}(2017)\citenamefont {Wu},
  \citenamefont {Liu}, \citenamefont {Yin},\ and\ \citenamefont
  {Lee}}]{Wu2017}%
  \BibitemOpen
  \bibfield  {author} {\bibinfo {author} {\bibfnamefont {B.}~\bibnamefont
  {Wu}}, \bibinfo {author} {\bibfnamefont {X.}~\bibnamefont {Liu}}, \bibinfo
  {author} {\bibfnamefont {J.}~\bibnamefont {Yin}},\ and\ \bibinfo {author}
  {\bibfnamefont {H.}~\bibnamefont {Lee}},\ }\bibfield  {title} {\bibinfo
  {title} {Bulk $\beta$-{Te} to few layered $\beta$-tellurenes: indirect to
  direct band-gap transitions showing semiconducting property},\ }\href
  {https://doi.org/10.1088/2053-1591/aa8ae3} {\bibfield  {journal} {\bibinfo
  {journal} {Mater. Res. Express}\ }\textbf {\bibinfo {volume} {4}},\ \bibinfo
  {pages} {095902} (\bibinfo {year} {2017})}\BibitemShut {NoStop}%
\bibitem [{\citenamefont {Niu}\ \emph {et~al.}(2023{\natexlab{a}})\citenamefont
  {Niu}, \citenamefont {Zhang}, \citenamefont {Graf}, \citenamefont {Lee},
  \citenamefont {Wang}, \citenamefont {Wu}, \citenamefont {Low},\ and\
  \citenamefont {Ye}}]{Niu2023b}%
  \BibitemOpen
  \bibfield  {author} {\bibinfo {author} {\bibfnamefont {C.}~\bibnamefont
  {Niu}}, \bibinfo {author} {\bibfnamefont {Z.}~\bibnamefont {Zhang}}, \bibinfo
  {author} {\bibfnamefont {D.}~\bibnamefont {Graf}}, \bibinfo {author}
  {\bibfnamefont {S.}~\bibnamefont {Lee}}, \bibinfo {author} {\bibfnamefont
  {M.}~\bibnamefont {Wang}}, \bibinfo {author} {\bibfnamefont {W.}~\bibnamefont
  {Wu}}, \bibinfo {author} {\bibfnamefont {T.}~\bibnamefont {Low}},\ and\
  \bibinfo {author} {\bibfnamefont {P.~D.}\ \bibnamefont {Ye}},\ }\bibfield
  {title} {\bibinfo {title} {High-pressure induced {Weyl} semimetal phase in 2d
  tellurium},\ }\href {https://doi.org/10.1038/s42005-023-01460-1} {\bibfield
  {journal} {\bibinfo  {journal} {Comm. Phys.}\ }\textbf {\bibinfo {volume}
  {6}},\ \bibinfo {pages} {345} (\bibinfo {year}
  {2023}{\natexlab{a}})}\BibitemShut {NoStop}%
\bibitem [{\citenamefont {Apte}\ \emph {et~al.}(2021)\citenamefont {Apte},
  \citenamefont {Kouser}, \citenamefont {Safi~Samghabadi}, \citenamefont
  {Chang}, \citenamefont {Sassi}, \citenamefont {Litvinov}, \citenamefont
  {Yakobson}, \citenamefont {Puthirath},\ and\ \citenamefont
  {Ajayan}}]{Apte2021}%
  \BibitemOpen
  \bibfield  {author} {\bibinfo {author} {\bibfnamefont {A.}~\bibnamefont
  {Apte}}, \bibinfo {author} {\bibfnamefont {S.}~\bibnamefont {Kouser}},
  \bibinfo {author} {\bibfnamefont {F.}~\bibnamefont {Safi~Samghabadi}},
  \bibinfo {author} {\bibfnamefont {L.}~\bibnamefont {Chang}}, \bibinfo
  {author} {\bibfnamefont {L.~M.}\ \bibnamefont {Sassi}}, \bibinfo {author}
  {\bibfnamefont {D.}~\bibnamefont {Litvinov}}, \bibinfo {author}
  {\bibfnamefont {B.~I.}\ \bibnamefont {Yakobson}}, \bibinfo {author}
  {\bibfnamefont {A.~B.}\ \bibnamefont {Puthirath}},\ and\ \bibinfo {author}
  {\bibfnamefont {P.~M.}\ \bibnamefont {Ajayan}},\ }\bibfield  {title}
  {\bibinfo {title} {Piezo-response in two-dimensional $\alpha$-tellurene
  films},\ }\href {https://doi.org/10.1016/j.mattod.2020.10.030} {\bibfield
  {journal} {\bibinfo  {journal} {Mater. Today}\ }\textbf {\bibinfo {volume}
  {44}},\ \bibinfo {pages} {40} (\bibinfo {year} {2021})}\BibitemShut {NoStop}%
\bibitem [{\citenamefont {Rao}\ \emph {et~al.}(2022)\citenamefont {Rao},
  \citenamefont {Fang}, \citenamefont {Zhou}, \citenamefont {Zhao},
  \citenamefont {Shang}, \citenamefont {Huang}, \citenamefont {Liu},
  \citenamefont {Du}, \citenamefont {Li}, \citenamefont {Jian}, \citenamefont
  {Ma}, \citenamefont {Wang}, \citenamefont {Liu}, \citenamefont {Wu},
  \citenamefont {Wang},\ and\ \citenamefont {Xiong}}]{Rao2022}%
  \BibitemOpen
  \bibfield  {author} {\bibinfo {author} {\bibfnamefont {G.}~\bibnamefont
  {Rao}}, \bibinfo {author} {\bibfnamefont {H.}~\bibnamefont {Fang}}, \bibinfo
  {author} {\bibfnamefont {T.}~\bibnamefont {Zhou}}, \bibinfo {author}
  {\bibfnamefont {C.}~\bibnamefont {Zhao}}, \bibinfo {author} {\bibfnamefont
  {N.}~\bibnamefont {Shang}}, \bibinfo {author} {\bibfnamefont
  {J.}~\bibnamefont {Huang}}, \bibinfo {author} {\bibfnamefont
  {Y.}~\bibnamefont {Liu}}, \bibinfo {author} {\bibfnamefont {X.}~\bibnamefont
  {Du}}, \bibinfo {author} {\bibfnamefont {P.}~\bibnamefont {Li}}, \bibinfo
  {author} {\bibfnamefont {X.}~\bibnamefont {Jian}}, \bibinfo {author}
  {\bibfnamefont {L.}~\bibnamefont {Ma}}, \bibinfo {author} {\bibfnamefont
  {J.}~\bibnamefont {Wang}}, \bibinfo {author} {\bibfnamefont {K.}~\bibnamefont
  {Liu}}, \bibinfo {author} {\bibfnamefont {J.}~\bibnamefont {Wu}}, \bibinfo
  {author} {\bibfnamefont {X.}~\bibnamefont {Wang}},\ and\ \bibinfo {author}
  {\bibfnamefont {J.}~\bibnamefont {Xiong}},\ }\bibfield  {title} {\bibinfo
  {title} {Robust piezoelectricity with spontaneous polarization in monolayer
  tellurene and multilayer tellurium film at room temperature for reliable
  memory},\ }\href {https://doi.org/10.1002/adma.202204697} {\bibfield
  {journal} {\bibinfo  {journal} {Adv. Mater.}\ }\textbf {\bibinfo {volume}
  {34}},\ \bibinfo {pages} {2204697} (\bibinfo {year} {2022})}\BibitemShut
  {NoStop}%
\bibitem [{\citenamefont {Gao}\ \emph {et~al.}(2019)\citenamefont {Gao},
  \citenamefont {Sun},\ and\ \citenamefont {Zhang}}]{Gao2019}%
  \BibitemOpen
  \bibfield  {author} {\bibinfo {author} {\bibfnamefont {S.}~\bibnamefont
  {Gao}}, \bibinfo {author} {\bibfnamefont {C.}~\bibnamefont {Sun}},\ and\
  \bibinfo {author} {\bibfnamefont {X.}~\bibnamefont {Zhang}},\ }\bibfield
  {title} {\bibinfo {title} {Ultra-strong anisotropic photo-responsivity of
  bilayer tellurene: a quantum transport and time-domain first principle
  study},\ }\href {https://doi.org/10.1515/nanoph-2019-0435} {\bibfield
  {journal} {\bibinfo  {journal} {Nanophotonics}\ }\textbf {\bibinfo {volume}
  {9}},\ \bibinfo {pages} {1931} (\bibinfo {year} {2019})}\BibitemShut
  {NoStop}%
\bibitem [{\citenamefont {Wang}\ \emph {et~al.}(2018)\citenamefont {Wang},
  \citenamefont {Qiu}, \citenamefont {Wang}, \citenamefont {Huang},
  \citenamefont {Wang}, \citenamefont {Liu}, \citenamefont {Du}, \citenamefont
  {Goddard}, \citenamefont {Kim}, \citenamefont {Xu}, \citenamefont {Ye},\ and\
  \citenamefont {Wu}}]{Wang2018}%
  \BibitemOpen
  \bibfield  {author} {\bibinfo {author} {\bibfnamefont {Y.}~\bibnamefont
  {Wang}}, \bibinfo {author} {\bibfnamefont {G.}~\bibnamefont {Qiu}}, \bibinfo
  {author} {\bibfnamefont {R.}~\bibnamefont {Wang}}, \bibinfo {author}
  {\bibfnamefont {S.}~\bibnamefont {Huang}}, \bibinfo {author} {\bibfnamefont
  {Q.}~\bibnamefont {Wang}}, \bibinfo {author} {\bibfnamefont {Y.}~\bibnamefont
  {Liu}}, \bibinfo {author} {\bibfnamefont {Y.}~\bibnamefont {Du}}, \bibinfo
  {author} {\bibfnamefont {W.~A.}\ \bibnamefont {Goddard}}, \bibinfo {author}
  {\bibfnamefont {M.~J.}\ \bibnamefont {Kim}}, \bibinfo {author} {\bibfnamefont
  {X.}~\bibnamefont {Xu}}, \bibinfo {author} {\bibfnamefont {P.~D.}\
  \bibnamefont {Ye}},\ and\ \bibinfo {author} {\bibfnamefont {W.}~\bibnamefont
  {Wu}},\ }\bibfield  {title} {\bibinfo {title} {Field-effect transistors made
  from solution-grown two-dimensional tellurene},\ }\href
  {https://doi.org/10.1038/s41928-018-0058-4} {\bibfield  {journal} {\bibinfo
  {journal} {Nature Electronics}\ }\textbf {\bibinfo {volume} {1}},\ \bibinfo
  {pages} {228} (\bibinfo {year} {2018})}\BibitemShut {NoStop}%
\bibitem [{\citenamefont {Amani}\ \emph {et~al.}(2018)\citenamefont {Amani},
  \citenamefont {Tan}, \citenamefont {Zhang}, \citenamefont {Zhao},
  \citenamefont {Bullock}, \citenamefont {Song}, \citenamefont {Kim},
  \citenamefont {Shrestha}, \citenamefont {Gao}, \citenamefont {Crozier},
  \citenamefont {Scott},\ and\ \citenamefont {Javey}}]{Amani2018}%
  \BibitemOpen
  \bibfield  {author} {\bibinfo {author} {\bibfnamefont {M.}~\bibnamefont
  {Amani}}, \bibinfo {author} {\bibfnamefont {C.}~\bibnamefont {Tan}}, \bibinfo
  {author} {\bibfnamefont {G.}~\bibnamefont {Zhang}}, \bibinfo {author}
  {\bibfnamefont {C.}~\bibnamefont {Zhao}}, \bibinfo {author} {\bibfnamefont
  {J.}~\bibnamefont {Bullock}}, \bibinfo {author} {\bibfnamefont
  {X.}~\bibnamefont {Song}}, \bibinfo {author} {\bibfnamefont {H.}~\bibnamefont
  {Kim}}, \bibinfo {author} {\bibfnamefont {V.~R.}\ \bibnamefont {Shrestha}},
  \bibinfo {author} {\bibfnamefont {Y.}~\bibnamefont {Gao}}, \bibinfo {author}
  {\bibfnamefont {K.~B.}\ \bibnamefont {Crozier}}, \bibinfo {author}
  {\bibfnamefont {M.}~\bibnamefont {Scott}},\ and\ \bibinfo {author}
  {\bibfnamefont {A.}~\bibnamefont {Javey}},\ }\bibfield  {title} {\bibinfo
  {title} {Solution-synthesized high-mobility tellurium nanoflakes for
  short-wave infrared photodetectors},\ }\href
  {https://doi.org/10.1021/acsnano.8b03424} {\bibfield  {journal} {\bibinfo
  {journal} {ACS Nano}\ }\textbf {\bibinfo {volume} {12}},\ \bibinfo {pages}
  {7253} (\bibinfo {year} {2018})}\BibitemShut {NoStop}%
\bibitem [{\citenamefont {Zhang}\ \emph {et~al.}(2020)\citenamefont {Zhang},
  \citenamefont {Wang}, \citenamefont {Xing}, \citenamefont {Man},
  \citenamefont {Zhang}, \citenamefont {Han}, \citenamefont {Zhang},\ and\
  \citenamefont {Fu}}]{Zhang2020}%
  \BibitemOpen
  \bibfield  {author} {\bibinfo {author} {\bibfnamefont {W.}~\bibnamefont
  {Zhang}}, \bibinfo {author} {\bibfnamefont {G.}~\bibnamefont {Wang}},
  \bibinfo {author} {\bibfnamefont {F.}~\bibnamefont {Xing}}, \bibinfo {author}
  {\bibfnamefont {Z.}~\bibnamefont {Man}}, \bibinfo {author} {\bibfnamefont
  {F.}~\bibnamefont {Zhang}}, \bibinfo {author} {\bibfnamefont
  {K.}~\bibnamefont {Han}}, \bibinfo {author} {\bibfnamefont {H.}~\bibnamefont
  {Zhang}},\ and\ \bibinfo {author} {\bibfnamefont {S.}~\bibnamefont {Fu}},\
  }\bibfield  {title} {\bibinfo {title} {Passively q-switched and mode-locked
  erbium-doped fiber lasers based on tellurene nanosheets as saturable
  absorber},\ }\href {https://doi.org/10.1364/oe.392944} {\bibfield  {journal}
  {\bibinfo  {journal} {Opt. Express}\ }\textbf {\bibinfo {volume} {28}},\
  \bibinfo {pages} {14729} (\bibinfo {year} {2020})}\BibitemShut {NoStop}%
\bibitem [{\citenamefont {Qiu}\ \emph {et~al.}(2020)\citenamefont {Qiu},
  \citenamefont {Niu}, \citenamefont {Wang}, \citenamefont {Si}, \citenamefont
  {Zhang}, \citenamefont {Wu},\ and\ \citenamefont {Ye}}]{Qiu2020}%
  \BibitemOpen
  \bibfield  {author} {\bibinfo {author} {\bibfnamefont {G.}~\bibnamefont
  {Qiu}}, \bibinfo {author} {\bibfnamefont {C.}~\bibnamefont {Niu}}, \bibinfo
  {author} {\bibfnamefont {Y.}~\bibnamefont {Wang}}, \bibinfo {author}
  {\bibfnamefont {M.}~\bibnamefont {Si}}, \bibinfo {author} {\bibfnamefont
  {Z.}~\bibnamefont {Zhang}}, \bibinfo {author} {\bibfnamefont
  {W.}~\bibnamefont {Wu}},\ and\ \bibinfo {author} {\bibfnamefont {P.~D.}\
  \bibnamefont {Ye}},\ }\bibfield  {title} {\bibinfo {title} {Quantum hall
  effect of weyl fermions in n-type semiconducting tellurene},\ }\href
  {https://doi.org/10.1038/s41565-020-0715-4} {\bibfield  {journal} {\bibinfo
  {journal} {Nat. Nanotechnology}\ }\textbf {\bibinfo {volume} {15}},\ \bibinfo
  {pages} {585} (\bibinfo {year} {2020})}\BibitemShut {NoStop}%
\bibitem [{\citenamefont {Niu}\ \emph {et~al.}(2021)\citenamefont {Niu},
  \citenamefont {Qiu}, \citenamefont {Wang}, \citenamefont {Si}, \citenamefont
  {Wu},\ and\ \citenamefont {Ye}}]{Niu2021}%
  \BibitemOpen
  \bibfield  {author} {\bibinfo {author} {\bibfnamefont {C.}~\bibnamefont
  {Niu}}, \bibinfo {author} {\bibfnamefont {G.}~\bibnamefont {Qiu}}, \bibinfo
  {author} {\bibfnamefont {Y.}~\bibnamefont {Wang}}, \bibinfo {author}
  {\bibfnamefont {M.}~\bibnamefont {Si}}, \bibinfo {author} {\bibfnamefont
  {W.}~\bibnamefont {Wu}},\ and\ \bibinfo {author} {\bibfnamefont {P.~D.}\
  \bibnamefont {Ye}},\ }\bibfield  {title} {\bibinfo {title} {Bilayer quantum
  hall states in an n-type wide tellurium quantum well},\ }\href
  {https://doi.org/10.1021/acs.nanolett.1c01705} {\bibfield  {journal}
  {\bibinfo  {journal} {Nano Lett.}\ }\textbf {\bibinfo {volume} {21}},\
  \bibinfo {pages} {7527} (\bibinfo {year} {2021})}\BibitemShut {NoStop}%
\bibitem [{\citenamefont {Sachdeva}\ \emph {et~al.}(2023)\citenamefont
  {Sachdeva}, \citenamefont {Gupta},\ and\ \citenamefont
  {Bera}}]{Sachdeva2023}%
  \BibitemOpen
  \bibfield  {author} {\bibinfo {author} {\bibfnamefont {P.~K.}\ \bibnamefont
  {Sachdeva}}, \bibinfo {author} {\bibfnamefont {S.}~\bibnamefont {Gupta}},\
  and\ \bibinfo {author} {\bibfnamefont {C.}~\bibnamefont {Bera}},\ }\bibfield
  {title} {\bibinfo {title} {Spin {Hall} effect induced by strain coupling of
  valley and spin polarization in puckered monochalcogenide tellurene
  monolayer},\ }\href {https://doi.org/10.1103/physrevb.107.155420} {\bibfield
  {journal} {\bibinfo  {journal} {Phys. Rev. B}\ }\textbf {\bibinfo {volume}
  {107}},\ \bibinfo {pages} {155420} (\bibinfo {year} {2023})}\BibitemShut
  {NoStop}%
\bibitem [{\citenamefont {Niu}\ \emph {et~al.}(2020)\citenamefont {Niu},
  \citenamefont {Qiu}, \citenamefont {Wang}, \citenamefont {Zhang},
  \citenamefont {Si}, \citenamefont {Wu},\ and\ \citenamefont {Ye}}]{Niu2020}%
  \BibitemOpen
  \bibfield  {author} {\bibinfo {author} {\bibfnamefont {C.}~\bibnamefont
  {Niu}}, \bibinfo {author} {\bibfnamefont {G.}~\bibnamefont {Qiu}}, \bibinfo
  {author} {\bibfnamefont {Y.}~\bibnamefont {Wang}}, \bibinfo {author}
  {\bibfnamefont {Z.}~\bibnamefont {Zhang}}, \bibinfo {author} {\bibfnamefont
  {M.}~\bibnamefont {Si}}, \bibinfo {author} {\bibfnamefont {W.}~\bibnamefont
  {Wu}},\ and\ \bibinfo {author} {\bibfnamefont {P.~D.}\ \bibnamefont {Ye}},\
  }\bibfield  {title} {\bibinfo {title} {Gate-tunable strong spin-orbit
  interaction in two-dimensional tellurium probed by weak antilocalization},\
  }\href {https://doi.org/10.1103/physrevb.101.205414} {\bibfield  {journal}
  {\bibinfo  {journal} {Phys. Rev. B}\ }\textbf {\bibinfo {volume} {101}},\
  \bibinfo {pages} {205414} (\bibinfo {year} {2020})}\BibitemShut {NoStop}%
\bibitem [{\citenamefont {Niu}\ \emph {et~al.}(2023{\natexlab{b}})\citenamefont
  {Niu}, \citenamefont {Qiu}, \citenamefont {Wang}, \citenamefont {Tan},
  \citenamefont {Wang}, \citenamefont {Jian}, \citenamefont {Wang},
  \citenamefont {Wu},\ and\ \citenamefont {Ye}}]{Niu2023}%
  \BibitemOpen
  \bibfield  {author} {\bibinfo {author} {\bibfnamefont {C.}~\bibnamefont
  {Niu}}, \bibinfo {author} {\bibfnamefont {G.}~\bibnamefont {Qiu}}, \bibinfo
  {author} {\bibfnamefont {Y.}~\bibnamefont {Wang}}, \bibinfo {author}
  {\bibfnamefont {P.}~\bibnamefont {Tan}}, \bibinfo {author} {\bibfnamefont
  {M.}~\bibnamefont {Wang}}, \bibinfo {author} {\bibfnamefont {J.}~\bibnamefont
  {Jian}}, \bibinfo {author} {\bibfnamefont {H.}~\bibnamefont {Wang}}, \bibinfo
  {author} {\bibfnamefont {W.}~\bibnamefont {Wu}},\ and\ \bibinfo {author}
  {\bibfnamefont {P.~D.}\ \bibnamefont {Ye}},\ }\bibfield  {title} {\bibinfo
  {title} {Tunable chirality-dependent nonlinear electrical responses in {2D}
  tellurium},\ }\href {https://doi.org/10.1021/acs.nanolett.3c01797} {\bibfield
   {journal} {\bibinfo  {journal} {Nano Lett.}\ }\textbf {\bibinfo {volume}
  {23}},\ \bibinfo {pages} {8445} (\bibinfo {year}
  {2023}{\natexlab{b}})}\BibitemShut {NoStop}%
\bibitem [{\citenamefont {Ma}\ \emph {et~al.}(2024)\citenamefont {Ma},
  \citenamefont {Xie}, \citenamefont {Zhang}, \citenamefont {Liu},
  \citenamefont {Ye}, \citenamefont {Yang}, \citenamefont {Liu}, \citenamefont
  {Chu}, \citenamefont {Liu}, \citenamefont {Zeng}, \citenamefont {Lu},\ and\
  \citenamefont {Wang}}]{Ma2024}%
  \BibitemOpen
  \bibfield  {author} {\bibinfo {author} {\bibfnamefont {B.}~\bibnamefont
  {Ma}}, \bibinfo {author} {\bibfnamefont {M.}~\bibnamefont {Xie}}, \bibinfo
  {author} {\bibfnamefont {L.}~\bibnamefont {Zhang}}, \bibinfo {author}
  {\bibfnamefont {S.}~\bibnamefont {Liu}}, \bibinfo {author} {\bibfnamefont
  {G.}~\bibnamefont {Ye}}, \bibinfo {author} {\bibfnamefont {X.}~\bibnamefont
  {Yang}}, \bibinfo {author} {\bibfnamefont {Y.}~\bibnamefont {Liu}}, \bibinfo
  {author} {\bibfnamefont {F.}~\bibnamefont {Chu}}, \bibinfo {author}
  {\bibfnamefont {Y.}~\bibnamefont {Liu}}, \bibinfo {author} {\bibfnamefont
  {X.}~\bibnamefont {Zeng}}, \bibinfo {author} {\bibfnamefont {X.}~\bibnamefont
  {Lu}},\ and\ \bibinfo {author} {\bibfnamefont {X.}~\bibnamefont {Wang}},\
  }\bibfield  {title} {\bibinfo {title} {Weyl node-participated
  magnetoresistance and nonreciprocal transport in two-dimensional tellurene
  nanostructures},\ }\href {https://doi.org/10.1021/acsanm.4c00510} {\bibfield
  {journal} {\bibinfo  {journal} {ACS Appl. Nano Mater.}\ }\textbf {\bibinfo
  {volume} {7}},\ \bibinfo {pages} {9012} (\bibinfo {year} {2024})}\BibitemShut
  {NoStop}%
\bibitem [{\citenamefont {Cheng}\ \emph {et~al.}(2024)\citenamefont {Cheng},
  \citenamefont {Gao}, \citenamefont {Zheng}, \citenamefont {Chen},
  \citenamefont {Liu}, \citenamefont {Zhang}, \citenamefont {Zhu},
  \citenamefont {Li}, \citenamefont {Li},\ and\ \citenamefont
  {Zeng}}]{Cheng2024}%
  \BibitemOpen
  \bibfield  {author} {\bibinfo {author} {\bibfnamefont {B.}~\bibnamefont
  {Cheng}}, \bibinfo {author} {\bibfnamefont {Y.}~\bibnamefont {Gao}}, \bibinfo
  {author} {\bibfnamefont {Z.}~\bibnamefont {Zheng}}, \bibinfo {author}
  {\bibfnamefont {S.}~\bibnamefont {Chen}}, \bibinfo {author} {\bibfnamefont
  {Z.}~\bibnamefont {Liu}}, \bibinfo {author} {\bibfnamefont {L.}~\bibnamefont
  {Zhang}}, \bibinfo {author} {\bibfnamefont {Q.}~\bibnamefont {Zhu}}, \bibinfo
  {author} {\bibfnamefont {H.}~\bibnamefont {Li}}, \bibinfo {author}
  {\bibfnamefont {L.}~\bibnamefont {Li}},\ and\ \bibinfo {author}
  {\bibfnamefont {C.}~\bibnamefont {Zeng}},\ }\bibfield  {title} {\bibinfo
  {title} {Giant nonlinear {Hall} and wireless rectification effects at room
  temperature in the elemental semiconductor tellurium},\ }\href
  {https://doi.org/10.1038/s41467-024-49706-y} {\bibfield  {journal} {\bibinfo
  {journal} {Nat. Commun.}\ }\textbf {\bibinfo {volume} {15}},\ \bibinfo
  {pages} {5513} (\bibinfo {year} {2024})}\BibitemShut {NoStop}%
\bibitem [{\citenamefont {Kim}\ \emph {et~al.}(2024)\citenamefont {Kim},
  \citenamefont {Bahng}, \citenamefont {Jeong}, \citenamefont {Sakong},
  \citenamefont {Lee}, \citenamefont {Lee}, \citenamefont {Kim}, \citenamefont
  {Rho},\ and\ \citenamefont {Lim}}]{Kim2024}%
  \BibitemOpen
  \bibfield  {author} {\bibinfo {author} {\bibfnamefont {G.}~\bibnamefont
  {Kim}}, \bibinfo {author} {\bibfnamefont {J.}~\bibnamefont {Bahng}}, \bibinfo
  {author} {\bibfnamefont {J.}~\bibnamefont {Jeong}}, \bibinfo {author}
  {\bibfnamefont {W.}~\bibnamefont {Sakong}}, \bibinfo {author} {\bibfnamefont
  {T.}~\bibnamefont {Lee}}, \bibinfo {author} {\bibfnamefont {D.}~\bibnamefont
  {Lee}}, \bibinfo {author} {\bibfnamefont {Y.}~\bibnamefont {Kim}}, \bibinfo
  {author} {\bibfnamefont {H.}~\bibnamefont {Rho}},\ and\ \bibinfo {author}
  {\bibfnamefont {S.~C.}\ \bibnamefont {Lim}},\ }\bibfield  {title} {\bibinfo
  {title} {Gate modulation of dissipationless nonlinear quantum geometric
  current in {2D Te}},\ }\href {https://doi.org/10.1021/acs.nanolett.4c02224}
  {\bibfield  {journal} {\bibinfo  {journal} {Nano Lett.}\ }\textbf {\bibinfo
  {volume} {24}},\ \bibinfo {pages} {10820} (\bibinfo {year}
  {2024})}\BibitemShut {NoStop}%
\bibitem [{\citenamefont {Sodemann}\ and\ \citenamefont
  {Fu}(2015)}]{Sodemann2015}%
  \BibitemOpen
  \bibfield  {author} {\bibinfo {author} {\bibfnamefont {I.}~\bibnamefont
  {Sodemann}}\ and\ \bibinfo {author} {\bibfnamefont {L.}~\bibnamefont {Fu}},\
  }\bibfield  {title} {\bibinfo {title} {Quantum nonlinear hall effect induced
  by berry curvature dipole in time-reversal invariant materials},\ }\href
  {https://doi.org/10.1103/physrevlett.115.216806} {\bibfield  {journal}
  {\bibinfo  {journal} {Phys. Rev. Lett.}\ }\textbf {\bibinfo {volume} {115}},\
  \bibinfo {pages} {216806} (\bibinfo {year} {2015})}\BibitemShut {NoStop}%
\bibitem [{\citenamefont {Ma}\ \emph {et~al.}(2018)\citenamefont {Ma},
  \citenamefont {Xu}, \citenamefont {Shen}, \citenamefont {MacNeill},
  \citenamefont {Fatemi}, \citenamefont {Chang}, \citenamefont {Mier~Valdivia},
  \citenamefont {Wu}, \citenamefont {Du}, \citenamefont {Hsu}, \citenamefont
  {Fang}, \citenamefont {Gibson}, \citenamefont {Watanabe}, \citenamefont
  {Taniguchi}, \citenamefont {Cava}, \citenamefont {Kaxiras}, \citenamefont
  {Lu}, \citenamefont {Lin}, \citenamefont {Fu}, \citenamefont {Gedik},\ and\
  \citenamefont {Jarillo-Herrero}}]{Ma2018a}%
  \BibitemOpen
  \bibfield  {author} {\bibinfo {author} {\bibfnamefont {Q.}~\bibnamefont
  {Ma}}, \bibinfo {author} {\bibfnamefont {S.-Y.}\ \bibnamefont {Xu}}, \bibinfo
  {author} {\bibfnamefont {H.}~\bibnamefont {Shen}}, \bibinfo {author}
  {\bibfnamefont {D.}~\bibnamefont {MacNeill}}, \bibinfo {author}
  {\bibfnamefont {V.}~\bibnamefont {Fatemi}}, \bibinfo {author} {\bibfnamefont
  {T.-R.}\ \bibnamefont {Chang}}, \bibinfo {author} {\bibfnamefont {A.~M.}\
  \bibnamefont {Mier~Valdivia}}, \bibinfo {author} {\bibfnamefont
  {S.}~\bibnamefont {Wu}}, \bibinfo {author} {\bibfnamefont {Z.}~\bibnamefont
  {Du}}, \bibinfo {author} {\bibfnamefont {C.-H.}\ \bibnamefont {Hsu}},
  \bibinfo {author} {\bibfnamefont {S.}~\bibnamefont {Fang}}, \bibinfo {author}
  {\bibfnamefont {Q.~D.}\ \bibnamefont {Gibson}}, \bibinfo {author}
  {\bibfnamefont {K.}~\bibnamefont {Watanabe}}, \bibinfo {author}
  {\bibfnamefont {T.}~\bibnamefont {Taniguchi}}, \bibinfo {author}
  {\bibfnamefont {R.~J.}\ \bibnamefont {Cava}}, \bibinfo {author}
  {\bibfnamefont {E.}~\bibnamefont {Kaxiras}}, \bibinfo {author} {\bibfnamefont
  {H.-Z.}\ \bibnamefont {Lu}}, \bibinfo {author} {\bibfnamefont
  {H.}~\bibnamefont {Lin}}, \bibinfo {author} {\bibfnamefont {L.}~\bibnamefont
  {Fu}}, \bibinfo {author} {\bibfnamefont {N.}~\bibnamefont {Gedik}},\ and\
  \bibinfo {author} {\bibfnamefont {P.}~\bibnamefont {Jarillo-Herrero}},\
  }\bibfield  {title} {\bibinfo {title} {Observation of the nonlinear hall
  effect under time-reversal-symmetric conditions},\ }\href
  {https://doi.org/10.1038/s41586-018-0807-6} {\bibfield  {journal} {\bibinfo
  {journal} {Nature}\ }\textbf {\bibinfo {volume} {565}},\ \bibinfo {pages}
  {337} (\bibinfo {year} {2018})}\BibitemShut {NoStop}%
\bibitem [{\citenamefont {Du}\ \emph {et~al.}(2021{\natexlab{a}})\citenamefont
  {Du}, \citenamefont {Lu},\ and\ \citenamefont {Xie}}]{Du2021}%
  \BibitemOpen
  \bibfield  {author} {\bibinfo {author} {\bibfnamefont {Z.~Z.}\ \bibnamefont
  {Du}}, \bibinfo {author} {\bibfnamefont {H.-Z.}\ \bibnamefont {Lu}},\ and\
  \bibinfo {author} {\bibfnamefont {X.~C.}\ \bibnamefont {Xie}},\ }\bibfield
  {title} {\bibinfo {title} {Nonlinear {Hall} effects},\ }\href
  {https://doi.org/10.1038/s42254-021-00359-6} {\bibfield  {journal} {\bibinfo
  {journal} {Nat. Rev. Phys.}\ }\textbf {\bibinfo {volume} {3}},\ \bibinfo
  {pages} {744} (\bibinfo {year} {2021}{\natexlab{a}})}\BibitemShut {NoStop}%
\bibitem [{\citenamefont {Du}\ \emph {et~al.}(2021{\natexlab{b}})\citenamefont
  {Du}, \citenamefont {Wang}, \citenamefont {Sun}, \citenamefont {Lu},\ and\
  \citenamefont {Xie}}]{Du2021a}%
  \BibitemOpen
  \bibfield  {author} {\bibinfo {author} {\bibfnamefont {Z.~Z.}\ \bibnamefont
  {Du}}, \bibinfo {author} {\bibfnamefont {C.~M.}\ \bibnamefont {Wang}},
  \bibinfo {author} {\bibfnamefont {H.-P.}\ \bibnamefont {Sun}}, \bibinfo
  {author} {\bibfnamefont {H.-Z.}\ \bibnamefont {Lu}},\ and\ \bibinfo {author}
  {\bibfnamefont {X.~C.}\ \bibnamefont {Xie}},\ }\bibfield  {title} {\bibinfo
  {title} {Quantum theory of the nonlinear hall effect},\ }\href
  {https://doi.org/10.1038/s41467-021-25273-4} {\bibfield  {journal} {\bibinfo
  {journal} {Nat. Commun.}\ }\textbf {\bibinfo {volume} {12}},\ \bibinfo
  {pages} {5038} (\bibinfo {year} {2021}{\natexlab{b}})}\BibitemShut {NoStop}%
\bibitem [{\citenamefont {Zhang}\ and\ \citenamefont {Fu}(2021)}]{Zhang2021a}%
  \BibitemOpen
  \bibfield  {author} {\bibinfo {author} {\bibfnamefont {Y.}~\bibnamefont
  {Zhang}}\ and\ \bibinfo {author} {\bibfnamefont {L.}~\bibnamefont {Fu}},\
  }\bibfield  {title} {\bibinfo {title} {Terahertz detection based on nonlinear
  {Hall} effect without magnetic field},\ }\href
  {https://doi.org/10.1073/pnas.2100736118} {\bibfield  {journal} {\bibinfo
  {journal} {Proc. Natl. Acad. Sci.}\ }\textbf {\bibinfo {volume} {118}},\
  \bibinfo {pages} {21} (\bibinfo {year} {2021})}\BibitemShut {NoStop}%
\bibitem [{\citenamefont {Ortix}(2021)}]{Ortix2021}%
  \BibitemOpen
  \bibfield  {author} {\bibinfo {author} {\bibfnamefont {C.}~\bibnamefont
  {Ortix}},\ }\bibfield  {title} {\bibinfo {title} {Nonlinear {Hall} effect
  with time‐reversal symmetry: {Theory} and material realizations},\ }\href
  {https://doi.org/10.1002/qute.202100056} {\bibfield  {journal} {\bibinfo
  {journal} {Adv. Quantum Technol.}\ }\textbf {\bibinfo {volume} {4}},\
  \bibinfo {pages} {2100056} (\bibinfo {year} {2021})}\BibitemShut {NoStop}%
\bibitem [{\citenamefont {Huang}\ \emph {et~al.}(2022)\citenamefont {Huang},
  \citenamefont {Wu}, \citenamefont {Hu}, \citenamefont {Cai}, \citenamefont
  {Li}, \citenamefont {An}, \citenamefont {Feng}, \citenamefont {Ye},
  \citenamefont {Lin}, \citenamefont {Law},\ and\ \citenamefont
  {Wang}}]{Huang2022a}%
  \BibitemOpen
  \bibfield  {author} {\bibinfo {author} {\bibfnamefont {M.}~\bibnamefont
  {Huang}}, \bibinfo {author} {\bibfnamefont {Z.}~\bibnamefont {Wu}}, \bibinfo
  {author} {\bibfnamefont {J.}~\bibnamefont {Hu}}, \bibinfo {author}
  {\bibfnamefont {X.}~\bibnamefont {Cai}}, \bibinfo {author} {\bibfnamefont
  {E.}~\bibnamefont {Li}}, \bibinfo {author} {\bibfnamefont {L.}~\bibnamefont
  {An}}, \bibinfo {author} {\bibfnamefont {X.}~\bibnamefont {Feng}}, \bibinfo
  {author} {\bibfnamefont {Z.}~\bibnamefont {Ye}}, \bibinfo {author}
  {\bibfnamefont {N.}~\bibnamefont {Lin}}, \bibinfo {author} {\bibfnamefont
  {K.~T.}\ \bibnamefont {Law}},\ and\ \bibinfo {author} {\bibfnamefont
  {N.}~\bibnamefont {Wang}},\ }\bibfield  {title} {\bibinfo {title} {Giant
  nonlinear hall effect in twisted bilayer wse2},\ }\href
  {https://doi.org/10.1093/nsr/nwac232} {\bibfield  {journal} {\bibinfo
  {journal} {Natl. Sci. Rev.}\ }\textbf {\bibinfo {volume} {10}},\ \bibinfo
  {pages} {4} (\bibinfo {year} {2022})}\BibitemShut {NoStop}%
\bibitem [{\citenamefont {Yar}\ \emph {et~al.}(2022)\citenamefont {Yar},
  \citenamefont {Jasra},\ and\ \citenamefont {Sabeeh}}]{Yar2022}%
  \BibitemOpen
  \bibfield  {author} {\bibinfo {author} {\bibfnamefont {A.}~\bibnamefont
  {Yar}}, \bibinfo {author} {\bibnamefont {Jasra}},\ and\ \bibinfo {author}
  {\bibfnamefont {K.}~\bibnamefont {Sabeeh}},\ }\bibfield  {title} {\bibinfo
  {title} {{Nonlinear {Hall} effect in topological insulator Bi$_2$Te$_3$ with
  hexagonal warping}},\ }\href {https://doi.org/10.1063/5.0090445} {\bibfield
  {journal} {\bibinfo  {journal} {J. Appl. Phys.}\ }\textbf {\bibinfo {volume}
  {131}},\ \bibinfo {pages} {184401} (\bibinfo {year} {2022})}\BibitemShut
  {NoStop}%
\bibitem [{\citenamefont {Gao}\ \emph {et~al.}(2023)\citenamefont {Gao},
  \citenamefont {Liu}, \citenamefont {Qiu}, \citenamefont {Ghosh},
  \citenamefont {V.~Trevisan}, \citenamefont {Onishi}, \citenamefont {Hu},
  \citenamefont {Qian}, \citenamefont {Tien}, \citenamefont {Chen},
  \citenamefont {Huang}, \citenamefont {Bérubé}, \citenamefont {Li},
  \citenamefont {Tzschaschel}, \citenamefont {Dinh}, \citenamefont {Sun},
  \citenamefont {Ho}, \citenamefont {Lien}, \citenamefont {Singh},
  \citenamefont {Watanabe}, \citenamefont {Taniguchi}, \citenamefont {Bell},
  \citenamefont {Lin}, \citenamefont {Chang}, \citenamefont {Du}, \citenamefont
  {Bansil}, \citenamefont {Fu}, \citenamefont {Ni}, \citenamefont {Orth},
  \citenamefont {Ma},\ and\ \citenamefont {Xu}}]{Gao2023}%
  \BibitemOpen
  \bibfield  {author} {\bibinfo {author} {\bibfnamefont {A.}~\bibnamefont
  {Gao}}, \bibinfo {author} {\bibfnamefont {Y.-F.}\ \bibnamefont {Liu}},
  \bibinfo {author} {\bibfnamefont {J.-X.}\ \bibnamefont {Qiu}}, \bibinfo
  {author} {\bibfnamefont {B.}~\bibnamefont {Ghosh}}, \bibinfo {author}
  {\bibfnamefont {T.}~\bibnamefont {V.~Trevisan}}, \bibinfo {author}
  {\bibfnamefont {Y.}~\bibnamefont {Onishi}}, \bibinfo {author} {\bibfnamefont
  {C.}~\bibnamefont {Hu}}, \bibinfo {author} {\bibfnamefont {T.}~\bibnamefont
  {Qian}}, \bibinfo {author} {\bibfnamefont {H.-J.}\ \bibnamefont {Tien}},
  \bibinfo {author} {\bibfnamefont {S.-W.}\ \bibnamefont {Chen}}, \bibinfo
  {author} {\bibfnamefont {M.}~\bibnamefont {Huang}}, \bibinfo {author}
  {\bibfnamefont {D.}~\bibnamefont {Bérubé}}, \bibinfo {author}
  {\bibfnamefont {H.}~\bibnamefont {Li}}, \bibinfo {author} {\bibfnamefont
  {C.}~\bibnamefont {Tzschaschel}}, \bibinfo {author} {\bibfnamefont
  {T.}~\bibnamefont {Dinh}}, \bibinfo {author} {\bibfnamefont {Z.}~\bibnamefont
  {Sun}}, \bibinfo {author} {\bibfnamefont {S.-C.}\ \bibnamefont {Ho}},
  \bibinfo {author} {\bibfnamefont {S.-W.}\ \bibnamefont {Lien}}, \bibinfo
  {author} {\bibfnamefont {B.}~\bibnamefont {Singh}}, \bibinfo {author}
  {\bibfnamefont {K.}~\bibnamefont {Watanabe}}, \bibinfo {author}
  {\bibfnamefont {T.}~\bibnamefont {Taniguchi}}, \bibinfo {author}
  {\bibfnamefont {D.~C.}\ \bibnamefont {Bell}}, \bibinfo {author}
  {\bibfnamefont {H.}~\bibnamefont {Lin}}, \bibinfo {author} {\bibfnamefont
  {T.-R.}\ \bibnamefont {Chang}}, \bibinfo {author} {\bibfnamefont {C.~R.}\
  \bibnamefont {Du}}, \bibinfo {author} {\bibfnamefont {A.}~\bibnamefont
  {Bansil}}, \bibinfo {author} {\bibfnamefont {L.}~\bibnamefont {Fu}}, \bibinfo
  {author} {\bibfnamefont {N.}~\bibnamefont {Ni}}, \bibinfo {author}
  {\bibfnamefont {P.~P.}\ \bibnamefont {Orth}}, \bibinfo {author}
  {\bibfnamefont {Q.}~\bibnamefont {Ma}},\ and\ \bibinfo {author}
  {\bibfnamefont {S.-Y.}\ \bibnamefont {Xu}},\ }\bibfield  {title} {\bibinfo
  {title} {Quantum metric nonlinear {Hall} effect in a topological
  antiferromagnetic heterostructure},\ }\href
  {https://doi.org/10.1126/science.adf1506} {\bibfield  {journal} {\bibinfo
  {journal} {Science}\ }\textbf {\bibinfo {volume} {381}},\ \bibinfo {pages}
  {181} (\bibinfo {year} {2023})}\BibitemShut {NoStop}%
\bibitem [{\citenamefont {Huang}\ \emph {et~al.}(2023)\citenamefont {Huang},
  \citenamefont {Wu}, \citenamefont {Zhang}, \citenamefont {Feng},
  \citenamefont {Zhou}, \citenamefont {Wang}, \citenamefont {Chen},
  \citenamefont {Cheng}, \citenamefont {Sun}, \citenamefont {Meng},\ and\
  \citenamefont {Wang}}]{Huang2023}%
  \BibitemOpen
  \bibfield  {author} {\bibinfo {author} {\bibfnamefont {M.}~\bibnamefont
  {Huang}}, \bibinfo {author} {\bibfnamefont {Z.}~\bibnamefont {Wu}}, \bibinfo
  {author} {\bibfnamefont {X.}~\bibnamefont {Zhang}}, \bibinfo {author}
  {\bibfnamefont {X.}~\bibnamefont {Feng}}, \bibinfo {author} {\bibfnamefont
  {Z.}~\bibnamefont {Zhou}}, \bibinfo {author} {\bibfnamefont {S.}~\bibnamefont
  {Wang}}, \bibinfo {author} {\bibfnamefont {Y.}~\bibnamefont {Chen}}, \bibinfo
  {author} {\bibfnamefont {C.}~\bibnamefont {Cheng}}, \bibinfo {author}
  {\bibfnamefont {K.}~\bibnamefont {Sun}}, \bibinfo {author} {\bibfnamefont
  {Z.~Y.}\ \bibnamefont {Meng}},\ and\ \bibinfo {author} {\bibfnamefont
  {N.}~\bibnamefont {Wang}},\ }\bibfield  {title} {\bibinfo {title} {Intrinsic
  nonlinear {Hall} effect and gate-switchable berry curvature sliding in
  twisted bilayer graphene},\ }\href
  {https://doi.org/10.1103/physrevlett.131.066301} {\bibfield  {journal}
  {\bibinfo  {journal} {Phys. Rev. Lett.}\ }\textbf {\bibinfo {volume} {131}},\
  \bibinfo {pages} {066301} (\bibinfo {year} {2023})}\BibitemShut {NoStop}%
\bibitem [{\citenamefont {Li}\ \emph {et~al.}(2023)\citenamefont {Li},
  \citenamefont {Li}, \citenamefont {Xiao}, \citenamefont {Liu}, \citenamefont
  {Wu}, \citenamefont {Gan}, \citenamefont {Han}, \citenamefont {Tang},
  \citenamefont {Zhang},\ and\ \citenamefont {Wang}}]{Li2023}%
  \BibitemOpen
  \bibfield  {author} {\bibinfo {author} {\bibfnamefont {H.}~\bibnamefont
  {Li}}, \bibinfo {author} {\bibfnamefont {M.}~\bibnamefont {Li}}, \bibinfo
  {author} {\bibfnamefont {R.-C.}\ \bibnamefont {Xiao}}, \bibinfo {author}
  {\bibfnamefont {W.}~\bibnamefont {Liu}}, \bibinfo {author} {\bibfnamefont
  {L.}~\bibnamefont {Wu}}, \bibinfo {author} {\bibfnamefont {W.}~\bibnamefont
  {Gan}}, \bibinfo {author} {\bibfnamefont {H.}~\bibnamefont {Han}}, \bibinfo
  {author} {\bibfnamefont {X.}~\bibnamefont {Tang}}, \bibinfo {author}
  {\bibfnamefont {C.}~\bibnamefont {Zhang}},\ and\ \bibinfo {author}
  {\bibfnamefont {J.}~\bibnamefont {Wang}},\ }\bibfield  {title} {\bibinfo
  {title} {Current induced second-order nonlinear {Hall} effect in bulk
  {WTe}$_2$},\ }\href {https://doi.org/10.1063/5.0172026} {\bibfield  {journal}
  {\bibinfo  {journal} {Appl. Phys. Lett.}\ }\textbf {\bibinfo {volume}
  {123}},\ \bibinfo {pages} {163102} (\bibinfo {year} {2023})}\BibitemShut
  {NoStop}%
\bibitem [{\citenamefont {Bandyopadhyay}\ \emph {et~al.}(2024)\citenamefont
  {Bandyopadhyay}, \citenamefont {Joseph},\ and\ \citenamefont
  {Narayan}}]{Bandyopadhyay2024}%
  \BibitemOpen
  \bibfield  {author} {\bibinfo {author} {\bibfnamefont {A.}~\bibnamefont
  {Bandyopadhyay}}, \bibinfo {author} {\bibfnamefont {N.~B.}\ \bibnamefont
  {Joseph}},\ and\ \bibinfo {author} {\bibfnamefont {A.}~\bibnamefont
  {Narayan}},\ }\bibfield  {title} {\bibinfo {title} {Non-linear {Hall}
  effects: {Mechanisms} and materials},\ }\href
  {https://doi.org/10.1016/j.mtelec.2024.100101} {\bibfield  {journal}
  {\bibinfo  {journal} {Materials Today Electronics}\ }\textbf {\bibinfo
  {volume} {8}},\ \bibinfo {pages} {100101} (\bibinfo {year}
  {2024})}\BibitemShut {NoStop}%
\bibitem [{\citenamefont {Lee}\ \emph {et~al.}(2024)\citenamefont {Lee},
  \citenamefont {Wang}, \citenamefont {Chen}, \citenamefont {Kwon},
  \citenamefont {Hwang}, \citenamefont {Cho}, \citenamefont {Son},
  \citenamefont {Han}, \citenamefont {Choi}, \citenamefont {Kim}, \citenamefont
  {Mo}, \citenamefont {Petrovic}, \citenamefont {Hwang}, \citenamefont {Park},
  \citenamefont {Jang},\ and\ \citenamefont {Ryu}}]{Lee2024}%
  \BibitemOpen
  \bibfield  {author} {\bibinfo {author} {\bibfnamefont {J.-E.}\ \bibnamefont
  {Lee}}, \bibinfo {author} {\bibfnamefont {A.}~\bibnamefont {Wang}}, \bibinfo
  {author} {\bibfnamefont {S.}~\bibnamefont {Chen}}, \bibinfo {author}
  {\bibfnamefont {M.}~\bibnamefont {Kwon}}, \bibinfo {author} {\bibfnamefont
  {J.}~\bibnamefont {Hwang}}, \bibinfo {author} {\bibfnamefont
  {M.}~\bibnamefont {Cho}}, \bibinfo {author} {\bibfnamefont {K.-H.}\
  \bibnamefont {Son}}, \bibinfo {author} {\bibfnamefont {D.-S.}\ \bibnamefont
  {Han}}, \bibinfo {author} {\bibfnamefont {J.~W.}\ \bibnamefont {Choi}},
  \bibinfo {author} {\bibfnamefont {Y.~D.}\ \bibnamefont {Kim}}, \bibinfo
  {author} {\bibfnamefont {S.-K.}\ \bibnamefont {Mo}}, \bibinfo {author}
  {\bibfnamefont {C.}~\bibnamefont {Petrovic}}, \bibinfo {author}
  {\bibfnamefont {C.}~\bibnamefont {Hwang}}, \bibinfo {author} {\bibfnamefont
  {S.~Y.}\ \bibnamefont {Park}}, \bibinfo {author} {\bibfnamefont
  {C.}~\bibnamefont {Jang}},\ and\ \bibinfo {author} {\bibfnamefont
  {H.}~\bibnamefont {Ryu}},\ }\bibfield  {title} {\bibinfo {title}
  {Spin-orbit-splitting-driven nonlinear {Hall} effect in {NbIrTe}$_4$},\
  }\href {https://doi.org/10.1038/s41467-024-47643-4} {\bibfield  {journal}
  {\bibinfo  {journal} {Nat. Commun.}\ }\textbf {\bibinfo {volume} {15}},\
  \bibinfo {pages} {3971} (\bibinfo {year} {2024})}\BibitemShut {NoStop}%
\bibitem [{\citenamefont {Asnin}\ \emph {et~al.}(1978)\citenamefont {Asnin},
  \citenamefont {Bakun}, \citenamefont {Danishevskii}, \citenamefont
  {Ivchenko}, \citenamefont {Pikus},\ and\ \citenamefont
  {Rogachev}}]{Asnin1978}%
  \BibitemOpen
  \bibfield  {author} {\bibinfo {author} {\bibfnamefont {V.~M.}\ \bibnamefont
  {Asnin}}, \bibinfo {author} {\bibfnamefont {A.~A.}\ \bibnamefont {Bakun}},
  \bibinfo {author} {\bibfnamefont {A.~M.}\ \bibnamefont {Danishevskii}},
  \bibinfo {author} {\bibfnamefont {E.~L.}\ \bibnamefont {Ivchenko}}, \bibinfo
  {author} {\bibfnamefont {G.~E.}\ \bibnamefont {Pikus}},\ and\ \bibinfo
  {author} {\bibfnamefont {A.~A.}\ \bibnamefont {Rogachev}},\ }\bibfield
  {title} {\bibinfo {title} {Observation of a photo-emf that depends on the
  sign of the circular-polarization of the light},\ }\href
  {http://jetpletters.ru/ps/1557/article_23830.shtml} {\bibfield  {journal}
  {\bibinfo  {journal} {JETP Lett.}\ }\textbf {\bibinfo {volume} {28}},\
  \bibinfo {pages} {74} (\bibinfo {year} {1978})}\BibitemShut {NoStop}%
\bibitem [{\citenamefont {Tsirkin}\ \emph {et~al.}(2018)\citenamefont
  {Tsirkin}, \citenamefont {Puente},\ and\ \citenamefont
  {Souza}}]{Tsirkin2018}%
  \BibitemOpen
  \bibfield  {author} {\bibinfo {author} {\bibfnamefont {S.~S.}\ \bibnamefont
  {Tsirkin}}, \bibinfo {author} {\bibfnamefont {P.~A.}\ \bibnamefont
  {Puente}},\ and\ \bibinfo {author} {\bibfnamefont {I.}~\bibnamefont
  {Souza}},\ }\bibfield  {title} {\bibinfo {title} {Gyrotropic effects in
  trigonal tellurium studied from first principles},\ }\href
  {https://doi.org/10.1103/physrevb.97.035158} {\bibfield  {journal} {\bibinfo
  {journal} {Phys. Rev. B}\ }\textbf {\bibinfo {volume} {97}},\ \bibinfo
  {pages} {035158} (\bibinfo {year} {2018})}\BibitemShut {NoStop}%
\bibitem [{\citenamefont {Moldavskaya}\ \emph {et~al.}(2023)\citenamefont
  {Moldavskaya}, \citenamefont {Golub}, \citenamefont {Danilov}, \citenamefont
  {Bel’kov}, \citenamefont {Weiss},\ and\ \citenamefont
  {Ganichev}}]{Moldavskaya2023}%
  \BibitemOpen
  \bibfield  {author} {\bibinfo {author} {\bibfnamefont {M.~D.}\ \bibnamefont
  {Moldavskaya}}, \bibinfo {author} {\bibfnamefont {L.~E.}\ \bibnamefont
  {Golub}}, \bibinfo {author} {\bibfnamefont {S.~N.}\ \bibnamefont {Danilov}},
  \bibinfo {author} {\bibfnamefont {V.~V.}\ \bibnamefont {Bel’kov}}, \bibinfo
  {author} {\bibfnamefont {D.}~\bibnamefont {Weiss}},\ and\ \bibinfo {author}
  {\bibfnamefont {S.~D.}\ \bibnamefont {Ganichev}},\ }\bibfield  {title}
  {\bibinfo {title} {Photocurrents in bulk tellurium},\ }\href
  {https://doi.org/10.1103/physrevb.108.235209} {\bibfield  {journal} {\bibinfo
   {journal} {Phys. Rev. B}\ }\textbf {\bibinfo {volume} {108}},\ \bibinfo
  {pages} {235209} (\bibinfo {year} {2023})}\BibitemShut {NoStop}%
\bibitem [{\citenamefont {Ivchenko}(2005)}]{Ivchenko2005}%
  \BibitemOpen
  \bibfield  {author} {\bibinfo {author} {\bibfnamefont {E.~L.}\ \bibnamefont
  {Ivchenko}},\ }\href@noop {} {\emph {\bibinfo {title} {Optical Spectroscopy
  of Semiconductor Nanostructures}}}\ (\bibinfo  {publisher} {Alpha Sci. Int.
  Ltd.},\ \bibinfo {address} {Harrow},\ \bibinfo {year} {2005})\BibitemShut
  {NoStop}%
\bibitem [{\citenamefont {Ganichev}\ and\ \citenamefont
  {Prettl}(2005)}]{Ganichev2005}%
  \BibitemOpen
  \bibfield  {author} {\bibinfo {author} {\bibfnamefont {S.~D.}\ \bibnamefont
  {Ganichev}}\ and\ \bibinfo {author} {\bibfnamefont {W.}~\bibnamefont
  {Prettl}},\ }\href
  {https://doi.org/10.1093/acprof:oso/9780198528302.001.0001} {\emph {\bibinfo
  {title} {Intense Terahertz Excitation of Semiconductors}}}\ (\bibinfo
  {publisher} {Oxford University Press},\ \bibinfo {address} {Oxford},\
  \bibinfo {year} {2005})\BibitemShut {NoStop}%
\bibitem [{\citenamefont {Ivchenko}\ and\ \citenamefont
  {Ganichev}(2018)}]{Ivchenko2018}%
  \BibitemOpen
  \bibfield  {author} {\bibinfo {author} {\bibfnamefont {E.~L.}\ \bibnamefont
  {Ivchenko}}\ and\ \bibinfo {author} {\bibfnamefont {S.~D.}\ \bibnamefont
  {Ganichev}},\ }\href@noop {} {\emph {\bibinfo {title} {Spin Physics in
  Semiconductors}}},\ edited by\ \bibinfo {editor} {\bibfnamefont {M.~I.}\
  \bibnamefont {Dyakonov}}\ (\bibinfo  {publisher} {Springer},\ \bibinfo {year}
  {2018})\ \bibinfo {note} {281-328}\BibitemShut {NoStop}%
\bibitem [{\citenamefont {Ganichev}(1999)}]{Ganichev1999}%
  \BibitemOpen
  \bibfield  {author} {\bibinfo {author} {\bibfnamefont {S.~D.}\ \bibnamefont
  {Ganichev}},\ }\bibfield  {title} {\bibinfo {title} {Tunnel ionization of
  deep impurities in semiconductors induced by terahertz electric fields},\
  }\href {https://doi.org/10.1016/s0921-4526(99)00637-7} {\bibfield  {journal}
  {\bibinfo  {journal} {Phys. B}\ }\textbf {\bibinfo {volume} {273-274}},\
  \bibinfo {pages} {737} (\bibinfo {year} {1999})}\BibitemShut {NoStop}%
\bibitem [{\citenamefont {Herrmann}\ \emph {et~al.}(2016)\citenamefont
  {Herrmann}, \citenamefont {Dmitriev}, \citenamefont {Kozlov}, \citenamefont
  {Schneider}, \citenamefont {Jentzsch}, \citenamefont {Kvon}, \citenamefont
  {Olbrich}, \citenamefont {Bel'kov}, \citenamefont {Bayer}, \citenamefont
  {Schuh}, \citenamefont {Bougeard}, \citenamefont {Kuczmik}, \citenamefont
  {Oltscher}, \citenamefont {Weiss},\ and\ \citenamefont
  {Ganichev}}]{Herrmann2016}%
  \BibitemOpen
  \bibfield  {author} {\bibinfo {author} {\bibfnamefont {T.}~\bibnamefont
  {Herrmann}}, \bibinfo {author} {\bibfnamefont {I.~A.}\ \bibnamefont
  {Dmitriev}}, \bibinfo {author} {\bibfnamefont {D.~A.}\ \bibnamefont
  {Kozlov}}, \bibinfo {author} {\bibfnamefont {M.}~\bibnamefont {Schneider}},
  \bibinfo {author} {\bibfnamefont {B.}~\bibnamefont {Jentzsch}}, \bibinfo
  {author} {\bibfnamefont {Z.~D.}\ \bibnamefont {Kvon}}, \bibinfo {author}
  {\bibfnamefont {P.}~\bibnamefont {Olbrich}}, \bibinfo {author} {\bibfnamefont
  {V.~V.}\ \bibnamefont {Bel'kov}}, \bibinfo {author} {\bibfnamefont
  {A.}~\bibnamefont {Bayer}}, \bibinfo {author} {\bibfnamefont
  {D.}~\bibnamefont {Schuh}}, \bibinfo {author} {\bibfnamefont
  {D.}~\bibnamefont {Bougeard}}, \bibinfo {author} {\bibfnamefont
  {T.}~\bibnamefont {Kuczmik}}, \bibinfo {author} {\bibfnamefont
  {M.}~\bibnamefont {Oltscher}}, \bibinfo {author} {\bibfnamefont
  {D.}~\bibnamefont {Weiss}},\ and\ \bibinfo {author} {\bibfnamefont {S.~D.}\
  \bibnamefont {Ganichev}},\ }\bibfield  {title} {\bibinfo {title} {Analog of
  microwave-induced resistance oscillations induced in {GaAs} heterostructures
  by terahertz radiation},\ }\href {https://doi.org/10.1103/physrevb.94.081301}
  {\bibfield  {journal} {\bibinfo  {journal} {Phys. Rev. B}\ }\textbf {\bibinfo
  {volume} {94}},\ \bibinfo {pages} {081301} (\bibinfo {year}
  {2016})}\BibitemShut {NoStop}%
\bibitem [{\citenamefont {Saleh}\ and\ \citenamefont
  {Teich}(2019)}]{Saleh2019}%
  \BibitemOpen
  \bibfield  {author} {\bibinfo {author} {\bibfnamefont {B.~E.~A.}\
  \bibnamefont {Saleh}}\ and\ \bibinfo {author} {\bibfnamefont {M.~C.}\
  \bibnamefont {Teich}},\ }\href@noop {} {\emph {\bibinfo {title} {Fundamentals
  of Photonics}}}\ (\bibinfo  {publisher} {John Wiley and Sons Ltd.},\ \bibinfo
  {year} {2019})\BibitemShut {NoStop}%
\bibitem [{\citenamefont {Bel'kov}\ \emph {et~al.}(2005)\citenamefont
  {Bel'kov}, \citenamefont {Ganichev}, \citenamefont {Ivchenko}, \citenamefont
  {Tarasenko}, \citenamefont {Weber}, \citenamefont {Giglberger}, \citenamefont
  {Olteanu}, \citenamefont {Tranitz}, \citenamefont {Danilov}, \citenamefont
  {Schneider}, \citenamefont {Wegscheider}, \citenamefont {Weiss},\ and\
  \citenamefont {Prettl}}]{Belkov2005}%
  \BibitemOpen
  \bibfield  {author} {\bibinfo {author} {\bibfnamefont {V.~V.}\ \bibnamefont
  {Bel'kov}}, \bibinfo {author} {\bibfnamefont {S.~D.}\ \bibnamefont
  {Ganichev}}, \bibinfo {author} {\bibfnamefont {E.~L.}\ \bibnamefont
  {Ivchenko}}, \bibinfo {author} {\bibfnamefont {S.~A.}\ \bibnamefont
  {Tarasenko}}, \bibinfo {author} {\bibfnamefont {W.}~\bibnamefont {Weber}},
  \bibinfo {author} {\bibfnamefont {S.}~\bibnamefont {Giglberger}}, \bibinfo
  {author} {\bibfnamefont {M.}~\bibnamefont {Olteanu}}, \bibinfo {author}
  {\bibfnamefont {H.~P.}\ \bibnamefont {Tranitz}}, \bibinfo {author}
  {\bibfnamefont {S.~N.}\ \bibnamefont {Danilov}}, \bibinfo {author}
  {\bibfnamefont {P.}~\bibnamefont {Schneider}}, \bibinfo {author}
  {\bibfnamefont {W.}~\bibnamefont {Wegscheider}}, \bibinfo {author}
  {\bibfnamefont {D.}~\bibnamefont {Weiss}},\ and\ \bibinfo {author}
  {\bibfnamefont {W.}~\bibnamefont {Prettl}},\ }\bibfield  {title} {\bibinfo
  {title} {Magneto-gyrotropic photogalvanic effects in semiconductor quantum
  wells},\ }\href {https://doi.org/10.1088/0953-8984/17/21/032} {\bibfield
  {journal} {\bibinfo  {journal} {J. Phys. Condens. Matter}\ }\textbf {\bibinfo
  {volume} {17}},\ \bibinfo {pages} {3405} (\bibinfo {year}
  {2005})}\BibitemShut {NoStop}%
\bibitem [{\citenamefont {Golub}\ \emph {et~al.}(2020)\citenamefont {Golub},
  \citenamefont {Ivchenko},\ and\ \citenamefont {Spivak}}]{Golub2020}%
  \BibitemOpen
  \bibfield  {author} {\bibinfo {author} {\bibfnamefont {L.~E.}\ \bibnamefont
  {Golub}}, \bibinfo {author} {\bibfnamefont {E.}~\bibnamefont {Ivchenko}},\
  and\ \bibinfo {author} {\bibfnamefont {B.}~\bibnamefont {Spivak}},\
  }\bibfield  {title} {\bibinfo {title} {Semiclassical theory of the circular
  photogalvanic effect in gyrotropic systems},\ }\href
  {https://doi.org/10.1103/physrevb.102.085202} {\bibfield  {journal} {\bibinfo
   {journal} {Phys. Rev. B}\ }\textbf {\bibinfo {volume} {102}},\ \bibinfo
  {pages} {085202} (\bibinfo {year} {2020})}\BibitemShut {NoStop}%
\bibitem [{\citenamefont {Otteneder}\ \emph {et~al.}(2020)\citenamefont
  {Otteneder}, \citenamefont {Hubmann}, \citenamefont {Lu}, \citenamefont
  {Kozlov}, \citenamefont {Golub}, \citenamefont {Watanabe}, \citenamefont
  {Taniguchi}, \citenamefont {Efetov},\ and\ \citenamefont
  {Ganichev}}]{Otteneder2020}%
  \BibitemOpen
  \bibfield  {author} {\bibinfo {author} {\bibfnamefont {M.}~\bibnamefont
  {Otteneder}}, \bibinfo {author} {\bibfnamefont {S.}~\bibnamefont {Hubmann}},
  \bibinfo {author} {\bibfnamefont {X.}~\bibnamefont {Lu}}, \bibinfo {author}
  {\bibfnamefont {D.~A.}\ \bibnamefont {Kozlov}}, \bibinfo {author}
  {\bibfnamefont {L.~E.}\ \bibnamefont {Golub}}, \bibinfo {author}
  {\bibfnamefont {K.}~\bibnamefont {Watanabe}}, \bibinfo {author}
  {\bibfnamefont {T.}~\bibnamefont {Taniguchi}}, \bibinfo {author}
  {\bibfnamefont {D.~K.}\ \bibnamefont {Efetov}},\ and\ \bibinfo {author}
  {\bibfnamefont {S.~D.}\ \bibnamefont {Ganichev}},\ }\bibfield  {title}
  {\bibinfo {title} {Terahertz photogalvanics in twisted bilayer graphene close
  to the second magic angle},\ }\href
  {https://doi.org/10.1021/acs.nanolett.0c02474} {\bibfield  {journal}
  {\bibinfo  {journal} {Nano Lett.}\ }\textbf {\bibinfo {volume} {20}},\
  \bibinfo {pages} {7152} (\bibinfo {year} {2020})}\BibitemShut {NoStop}%
\bibitem [{Note1()}]{Note1}%
  \BibitemOpen
  \bibinfo {note} {Note that the circular photocurrent (CPGE) excited by
  infrared radiation with photon energy exceeding the energy gap and caused by
  direct interband optical transitions has been observed in Ref.~\cite
  {Niu2023a}.}\BibitemShut {Stop}%
\bibitem [{Note2()}]{Note2}%
  \BibitemOpen
  \bibinfo {note} {Note that for bulk tellurium the BCD contribution to the
  circular photocurrent along $C_3$-axis was calculated in Ref.~\cite
  {Tsirkin2018}.}\BibitemShut {Stop}%
\bibitem [{\citenamefont {Glazov}\ and\ \citenamefont
  {Golub}(2020{\natexlab{a}})}]{Glazov2020a}%
  \BibitemOpen
  \bibfield  {author} {\bibinfo {author} {\bibfnamefont {M.~M.}\ \bibnamefont
  {Glazov}}\ and\ \bibinfo {author} {\bibfnamefont {L.~E.}\ \bibnamefont
  {Golub}},\ }\bibfield  {title} {\bibinfo {title} {Skew scattering and side
  jump drive exciton valley {Hall} effect in two-dimensional crystals},\ }\href
  {https://doi.org/10.1103/PhysRevLett.125.157403} {\bibfield  {journal}
  {\bibinfo  {journal} {Phys. Rev. Lett.}\ }\textbf {\bibinfo {volume} {125}},\
  \bibinfo {pages} {157403} (\bibinfo {year} {2020}{\natexlab{a}})}\BibitemShut
  {NoStop}%
\bibitem [{\citenamefont {Glazov}\ and\ \citenamefont
  {Golub}(2020{\natexlab{b}})}]{Glazov2020}%
  \BibitemOpen
  \bibfield  {author} {\bibinfo {author} {\bibfnamefont {M.~M.}\ \bibnamefont
  {Glazov}}\ and\ \bibinfo {author} {\bibfnamefont {L.~E.}\ \bibnamefont
  {Golub}},\ }\bibfield  {title} {\bibinfo {title} {Valley hall effect caused
  by the phonon and photon drag},\ }\href
  {https://doi.org/10.1103/PhysRevB.102.155302} {\bibfield  {journal} {\bibinfo
   {journal} {Phys. Rev. B}\ }\textbf {\bibinfo {volume} {102}},\ \bibinfo
  {pages} {155302} (\bibinfo {year} {2020}{\natexlab{b}})}\BibitemShut
  {NoStop}%
\bibitem [{\citenamefont {Niu}\ \emph {et~al.}(2023{\natexlab{c}})\citenamefont
  {Niu}, \citenamefont {Huang}, \citenamefont {Ghosh}, \citenamefont {Tan},
  \citenamefont {Wang}, \citenamefont {Wu}, \citenamefont {Xu},\ and\
  \citenamefont {Ye}}]{Niu2023a}%
  \BibitemOpen
  \bibfield  {author} {\bibinfo {author} {\bibfnamefont {C.}~\bibnamefont
  {Niu}}, \bibinfo {author} {\bibfnamefont {S.}~\bibnamefont {Huang}}, \bibinfo
  {author} {\bibfnamefont {N.}~\bibnamefont {Ghosh}}, \bibinfo {author}
  {\bibfnamefont {P.}~\bibnamefont {Tan}}, \bibinfo {author} {\bibfnamefont
  {M.}~\bibnamefont {Wang}}, \bibinfo {author} {\bibfnamefont {W.}~\bibnamefont
  {Wu}}, \bibinfo {author} {\bibfnamefont {X.}~\bibnamefont {Xu}},\ and\
  \bibinfo {author} {\bibfnamefont {P.~D.}\ \bibnamefont {Ye}},\ }\bibfield
  {title} {\bibinfo {title} {Tunable circular photogalvanic and photovoltaic
  effect in {2D} tellurium with different chirality},\ }\href
  {https://doi.org/10.1021/acs.nanolett.3c00780} {\bibfield  {journal}
  {\bibinfo  {journal} {Nano Lett.}\ }\textbf {\bibinfo {volume} {23}},\
  \bibinfo {pages} {3599} (\bibinfo {year} {2023}{\natexlab{c}})}\BibitemShut
  {NoStop}%
\end{thebibliography}%

\end{document}